\documentclass[aps,10pt,twocolumn,showpacs,nofootinbib]{revtex4-1}

\usepackage{amsmath}
\usepackage{amsfonts}
\usepackage{amssymb}
\usepackage[pdftex]{graphicx}

\newcommand{\be}{\begin{equation}}
\newcommand{\ee}{\end{equation}}
\newcommand{\ba}{\begin{eqnarray}}
\newcommand{\ea}{\end{eqnarray}}

\def\g{\gamma}
\def\h{\eta}

\def\l{\lambda}
\def\m{\mu}

\def\r{\rho}

\def\D{\Delta}

\newcommand{\ov}{\overline}

\newcommand{\aand}{\;\;\;\mbox{and}\;\;\;}

\def\I{\leavevmode\hbox{\small1\kern-3.8pt\norfeynmpmalsize1}}

\begin{document}
\title{Gas diffusion among bubbles and the DCS risk}
 
\begin{abstract}
We present some experimental and simulation results that reproduces the Ostwald ripening (gas diffusion among bubbles) for air bubbles in a liquid fluid. Concerning the experiment, there it is measured the time evolution of bubbles mean radius, number of bubbles and radius size distribution. One of the main results shows that, while the number of bubbles decreases in time the bubbles mean radius increases, hence, it follows that the smaller bubbles disappear whereas the -- potentially dangerous for the diver -- larger bubbles grow up.  
Consequently, this effect suggests a possible  contribution of the Ostwald ripening to the decompression sickness, and if so, it should be pursued its implementation to the Reduced Gradient Bubble Model (RGBM) so as to build up dive tables and computer programs for further diving tests.
\end{abstract}

\author{Oswaldo M. Del Cima}
\email{oswaldo.delcima@ufv.br (NAUI Instructor $\sharp$48926)}
\author{Paulo C. Oliveira }
\author{C\'esar M. Rocha}
\author{Hallan S. Silva}
\author{Alvaro V.N.C. Teixeira}

\affiliation{Universidade Federal de Vi\c cosa (UFV),\\
Departamento de F\'\i sica - Campus Universit\'ario,\\
Avenida Peter Henry Rolfs s/n - 36570-900 -
Vi\c cosa - MG - Brazil.}

\maketitle

\centerline{In memory of Randy Shaw and Sergio Viegas} 

\section{Introduction}

The formation of gas bubbles \cite{diving_physics} -- due to nucleation (homogeneous or heterogeneous) and tribonucleation, and their evolution (expansion or contraction), owing to decompression or compression, diffusion, counterdiffusion, coalescence and Ostwald ripening -- in the blood and tissues of the human body can give rise to the decompression sickness (DCS). 
The Ostwald ripening mechanism consists in gas transfer from smaller bubbles to larger bubbles by diffusion in the liquid medium, consequently, the radii of larger bubbles increase at the expenses of decreasing radii of the smaller ones. 
It shall be presented the results of experiment and simulation which the Ostwald ripening is investigated for the case of gas (air) bubbles in a liquid fluid with some rheological parameters of the human blood. There, it has been measured and analyzed the time evolution of the bubbles mean radius, the number of bubbles and the radius size (frequency) distribution. At a fixed ambient pressure, namely, at the same ``depth'', one of the main experimental results has been undoubtedly shown that, while the number of bubbles decreases in time the bubbles mean radius increases, meaning that the smaller bubbles disappear whereas the larger (potentially dangerous) bubbles grow up. This phenomenon may reveal a contribution of the Ostwald ripening effect to the decompression sickness risk during and after diving, suggesting, therefore, a deeper theoretical and experimental investigation. Beyond that, if the Ostwald ripening shows up as an important physiological effect, its implementation to the RGBM (Reduced Gradient Bubble Model) \cite{diving_physics,rgbm,blood} for further diving tests should be pursued.

The outline of this presentation is as follows. In Section \ref{formation-macro} and Section \ref{formation-micro}, there are introduced the macroscopic and microscopic processes of bubble formation, respectively. The processes of bubble evolution are presented in Section \ref{evolution}, and the physiological consequences of 
bubble formation and evolution -- the decompression sickness -- are introduced in Section \ref{dcs1&2}. Section \ref{ostwald} introduces and discusses theoretical aspects of the Ostwald ripening phenomenon for gas bubbles in a liquid. The experiment, its apparatus and the result analysis are presented in Section \ref{experiment}. In Section \ref{simulation}, some preliminary results of the finite element simulation for one bubble, three, five and fifty bubbles are introduced. The conclusions and perspectives are left to Section \ref{conclusions}.  

\section{Macroscopic mechanisms of bubble formation}
\label{formation-macro}
\subsection{Cavitation}
\begin{figure}[h!]
\centering
\setlength{\unitlength}{1,0mm}
\includegraphics[width=4.25cm,height=2.65cm]{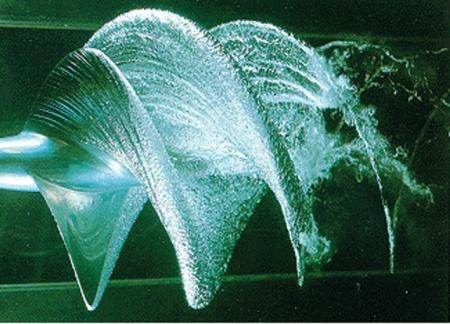}
\includegraphics[width=4.25cm,height=2.65cm]{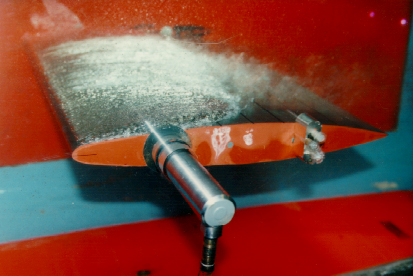}
\caption[]{Cavitation \cite{cavitation}.}
\label{cavitation}
\end{figure}

Cavitation is the process of rupturing a liquid by decreasing the pressure at roughly constant temperature -- quasi-isothermal process (FIG.\ref{cavitation}). 

\subsection{Boiling}
\begin{figure}[h!]
\centering
\setlength{\unitlength}{1,0mm}
\includegraphics[width=8.6cm,height=2.65cm]{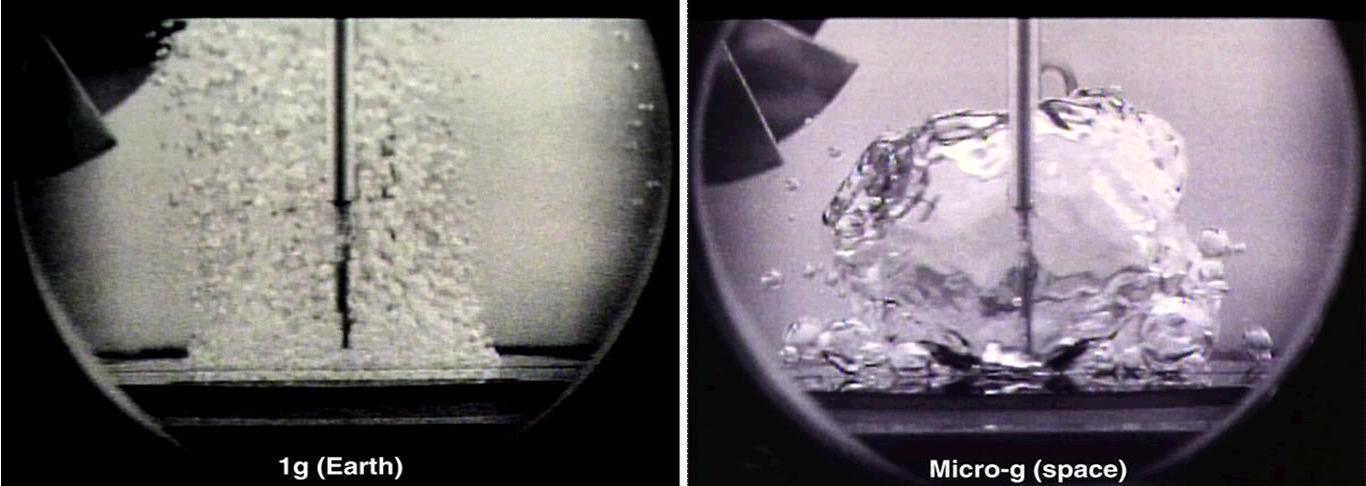}
\caption[]{Boiling \cite{boiling}.}
\label{boiling}
\end{figure}

Boiling is the process of rupturing a liquid by increasing the temperature at roughly constant pressure -- quasi-isobaric process (FIG.\ref{boiling}).

\section{Microscopic mechanisms of bubble formation}
\label{formation-micro}
\subsection{Nucleation}
Nucleation is a process of stochastic nature (microscopic fluctuations) that initiates the formation of new 
phase or structure.

\subsubsection{Homogeneous}
\begin{figure}[h!]
\centering
\setlength{\unitlength}{1,0mm}
\includegraphics[width=4.25cm,height=2.65cm,clip]{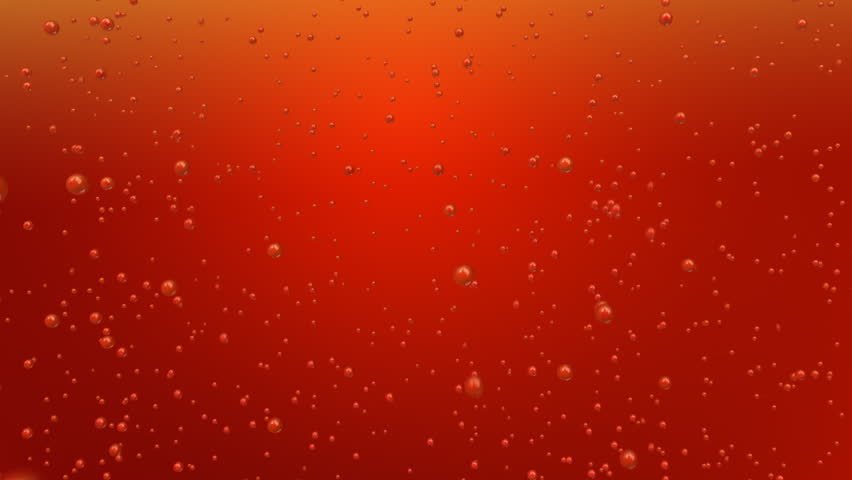}
\includegraphics[width=4.25cm,height=2.65cm,clip]{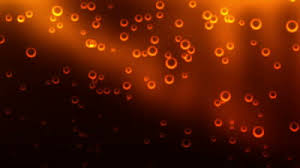}
\caption[]{Homogeneous nucleation \cite{homogeneous}.}
\label{homogeneous}
\end{figure}

Bubbles nucleate inside the bulk ­phase of a gas, liquid or solid (FIG.\ref{homogeneous}).

\subsubsection{Heterogeneous}
\begin{figure}[h!]
\centering
\setlength{\unitlength}{1,0mm}
\includegraphics[width=3.2cm,height=2.65cm,clip]{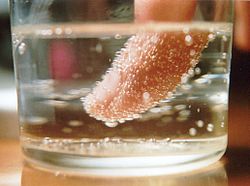}
\includegraphics[width=5.3cm,height=2.65cm,clip]{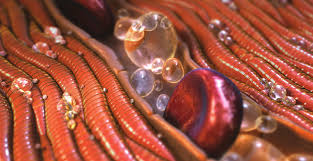}
\caption[]{Heterogeneous nucleation \cite{heterogeneous}.}
\label{heterogeneous}
\end{figure}

Bubbles nucleate upon liquid-­solid, gas-­liquid or gas-­solid, interfaces (FIG.\ref{heterogeneous}).

\subsection{Tribonucleation}
\begin{figure}[h!]
\centering
\setlength{\unitlength}{1,0mm}
\includegraphics[width=4.25cm,height=2.65cm,clip]{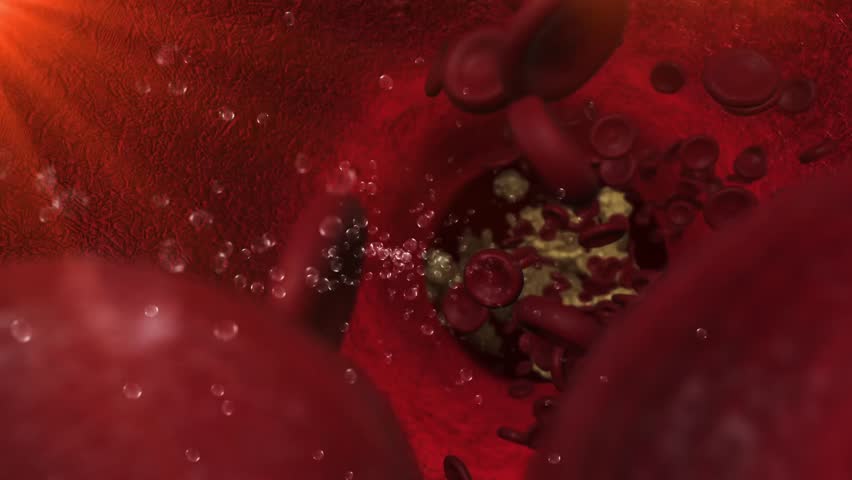}
\includegraphics[width=4.25cm,height=2.65cm,clip]{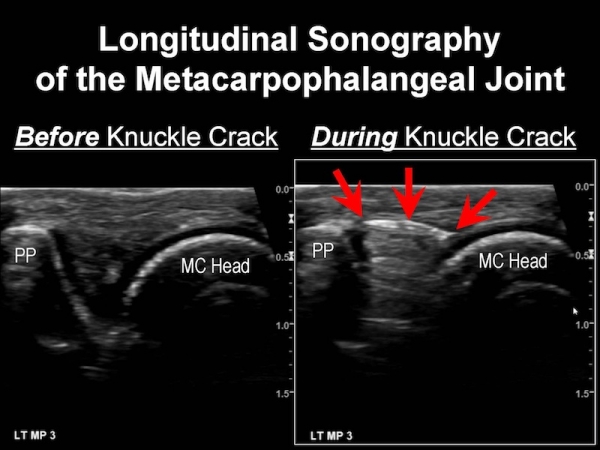}
\caption[]{Tribonucleation \cite{tribonucleation}.}
\label{tribonucleation}
\end{figure}

Tribonucleation is a gas microbubble formation process due to the relative movement among the liquid, containing dissolved gas, and a solid surface (FIG.\ref{tribonucleation}).

\subsubsection*{\underline{Bubble Formation}}
\begin{equation}
 \left\{
                \begin{array}{ll}
                  {\rm nucleation} \rightarrow \left\{
                \begin{array}{ll}
                  {\rm homogeneous} \\
                  {\rm heterogeneous}
                \end{array}
              \right.\\
              \\
                  {\rm tribonucleation}
                \end{array}
              \right. \nonumber
\end{equation}

\section{Mechanisms of bubble evolution}
\label{evolution}
\subsection{Decompression or compression}
\begin{figure}[h!]
\centering
\setlength{\unitlength}{1,0mm}
\includegraphics[width=3.2cm,height=2.65cm,clip]{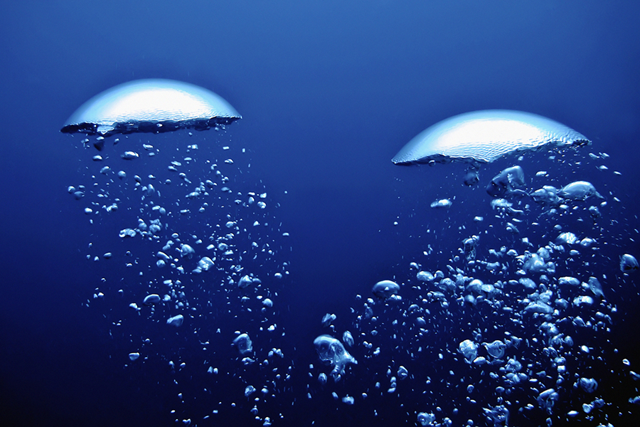}
\includegraphics[width=5.3cm,height=2.65cm,clip]{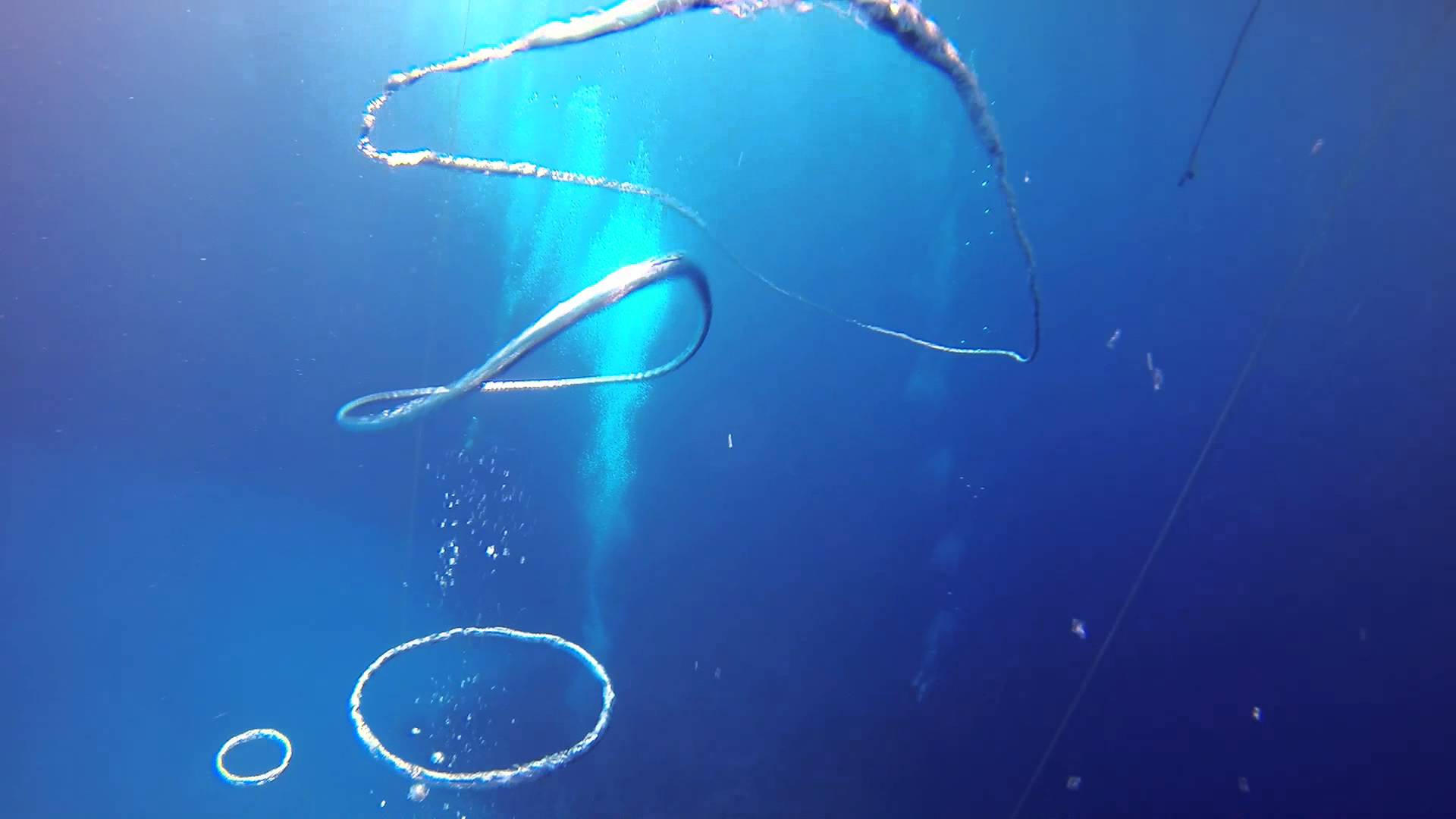}
\caption[]{Bubbles decompression \cite{decompression}.}
\label{decompression}
\end{figure}

Decompression (compression) is the process of decreasing (increasing) of the ambient pressure (FIG.\ref{decompression}).

\subsection{Coalescence}
\begin{figure}[h!]
\centering
\setlength{\unitlength}{1,0mm}
\includegraphics[width=3.2cm,height=2.65cm,clip]{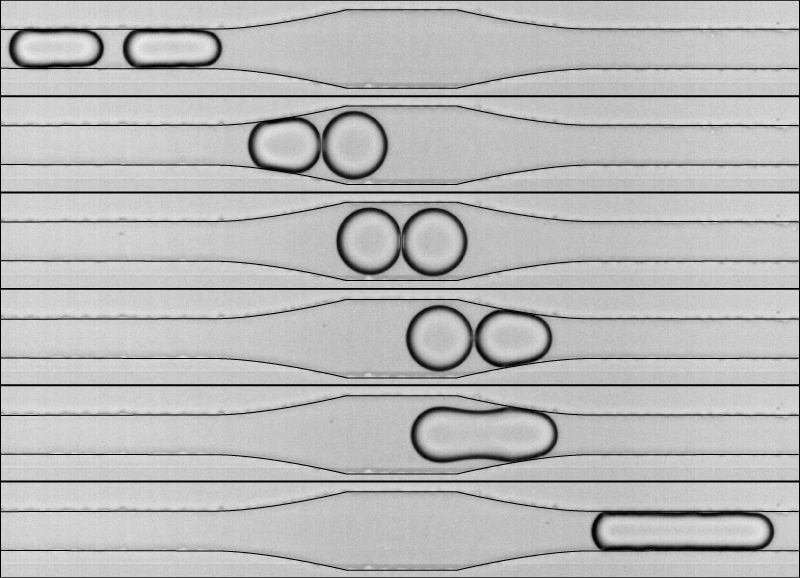}
\includegraphics[width=5.3cm,height=2.65cm,clip]{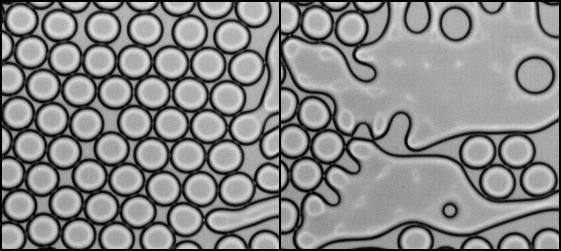}
\caption[]{Bubbles coalescence \cite{coalescence}.}
\label{coalescence}
\end{figure}

Coalescence is the fusion process of two or more bubbles. (FIG.\ref{coalescence}).

\subsection{Diffusion}
\begin{figure}[h!]
\centering
\setlength{\unitlength}{1,0mm}
\includegraphics[width=5.3cm,height=2.65cm,clip]{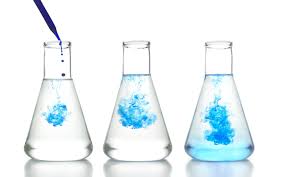}
\includegraphics[width=2.8cm,height=2.65cm,clip]{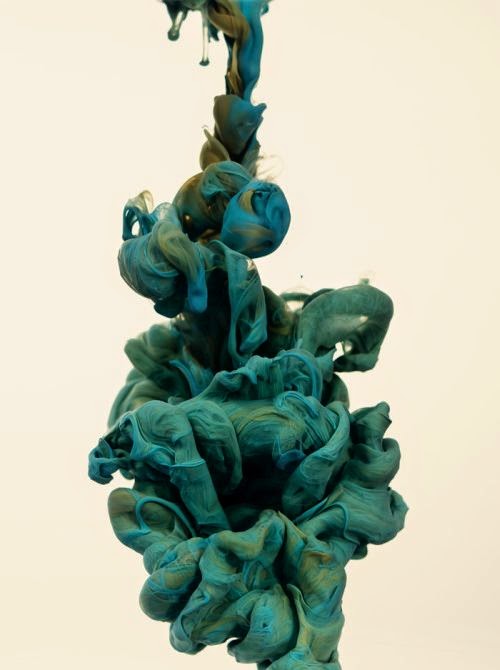}
\caption[]{Diffusion \cite{diffusion}.}
\label{diffusion}
\end{figure}

Diffusion is the flow of substance (atoms or molecules) from regions of higher concentration to regions of 
lower concentration (FIG.\ref{diffusion}).

\subsection{Isobaric counterdiffusion}
Isobaric counterdiffusion is the diffusion of different gases into and out of 
tissues while under a constant ambient pressure \cite{lambertsen-idicula,wienke}. When in diving, with multiple inert gases, and performing an isobaric gas mix switch, the inert components of the initial mix breathed by the diver begin to off-­gas the tissues, whereas the inert components of the second mix begin to in-­gas the tissues. ``There is no change in pressure and the gases are moving in opposite directions, this is called {\it isobaric counterdiffusion}'' \cite{taylor}.

\subsection{Ostwald ripening}
\begin{figure}[h!]
\centering
\setlength{\unitlength}{1,0mm}
\includegraphics[width=4.25cm,height=2.65cm,clip]{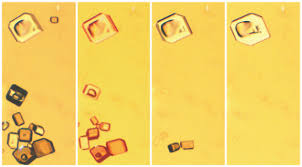}
\includegraphics[width=4.25cm,height=2.75cm,clip]{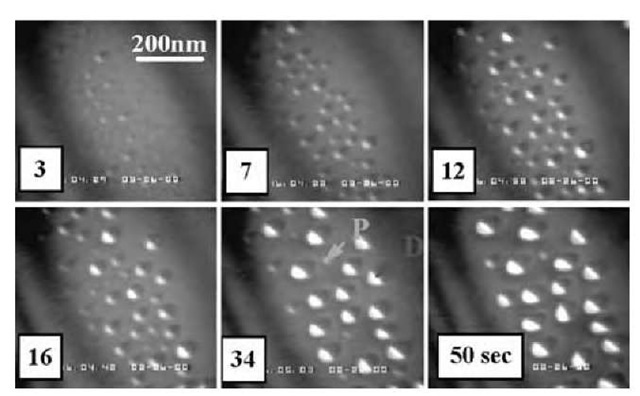}
\caption[]{Ostwald ripening: crystals and sol particles \cite{ostwald-crystals-sols}.}
\label{ostwald-crystals-sols}
\end{figure}

\begin{figure}[h!]
\centering
\setlength{\unitlength}{1,0mm}
\includegraphics[width=6.7cm,height=2.65cm]{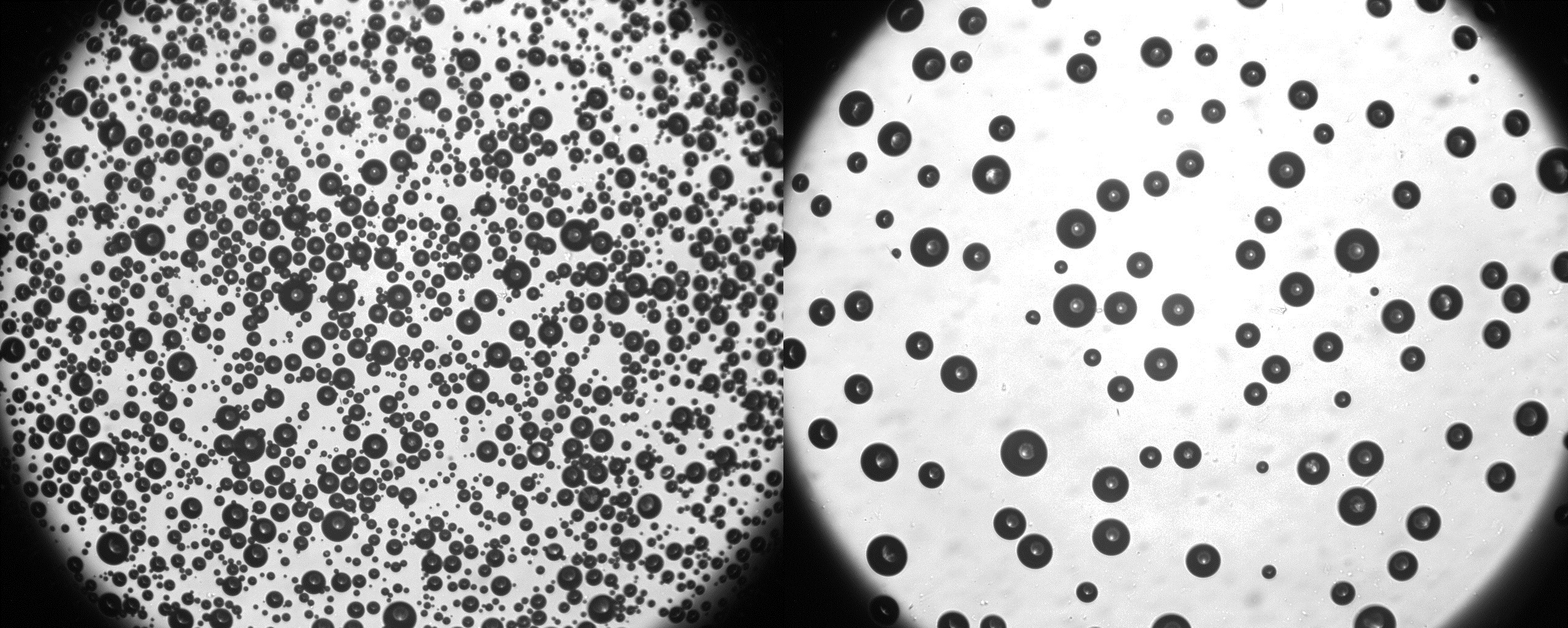}
\caption[]{Ostwald ripening: gas bubbles in a liquid fluid.}
\label{ostwald-gas-bubbles}
\end{figure}

In 1896, Wilhelm Ostwald has verified a phenomenon that small crystals (sol particles) dissolve and redeposit onto larger crystals (sol particles) \cite{ostwald}, which is called Ostwald ripening (FIG.\ref{ostwald-crystals-sols}). Later on, it has been already observed the Ostwald ripening among gas bubbles in liquid fluids, namely, gas transfer from smaller bubbles to larger bubbles (FIG.\ref{ostwald-gas-bubbles}).    

\subsubsection*{\underline{Bubble Evolution}}
\begin{equation}
 \left\{
        \begin{array}{ll}
                  {\rm present~on~\mathit{deco}~models} \rightarrow \left\{
           \begin{array}{ll}
                  {\rm decompression/compression} \\
                  {\rm diffusion} \\
                  {\rm isobaric~counterdiffusion}
           \end{array}
                \right. \\
                \\
             \begin{array}{ll}
                  {\rm absent~on~\mathit{deco}~models} \rightarrow \left\{
                 \begin{array}{ll}
                  {\rm coalescence} \\
                  {\rm \mathit{Ostwald~ripening}}
                 \end{array}
                 \right.
             \end{array}
       \end{array}
              \right. \nonumber
\end{equation}

\section{Decompression sickness}
\label{dcs1&2}
\begin{figure}[h!]
\centering
\setlength{\unitlength}{1,0mm}
\includegraphics[width=4.25cm,height=2.65cm]{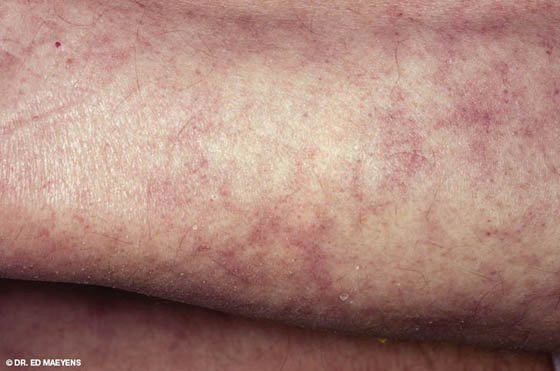}
\includegraphics[width=4.25cm,height=2.65cm]{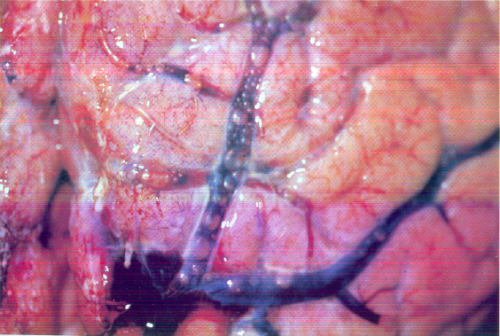}
\caption[]{Gas bubbles injuries: decompression sickness \cite{decompression-sickness}.}
\label{dcs}
\end{figure}

One of the hazards that divers, astronauts, aviators and compressed air workers, are subjected while under hyperbaric (or hypobaric) conditions, submitted to compression and decompression, is the decompression sickness (DCS) \cite{dcs}. These gas bubbles injuries, that trigger the decompression sickness, are due to the formation and evolution of intravascular and extravascular gas (N$_2$, He, O$_2$, CO$_2$, H$_2$O vapour) bubbles (FIG.\ref{dcs}). The collective insult of the gas bubbles to the body shall produce primary effects to the tissues which are directly insulted, further, the secondary effects can jeopardize the function of a wide range of tissues, therefore, compromising body's health, may even lead to its death. Decompression sickness is recognized by means of the signs and the symptoms exhibited by the body, just as its classification: Type 1  and Type 2. Type 1 DCS are usually characterized by mild cutaneous or skin symptoms, and musculoskeletal pain. Type 2 DCS symptoms are more severe, and they are typically split in three categories: cardiopulmonary, inner ear and neurological. 

\section{Ostwald ripening: gas bubbles in a liquid fluid}
\label{ostwald}
\begin{figure}[h!]
\centering
\setlength{\unitlength}{1,0mm}
\includegraphics[width=7.4cm,height=2.65cm]{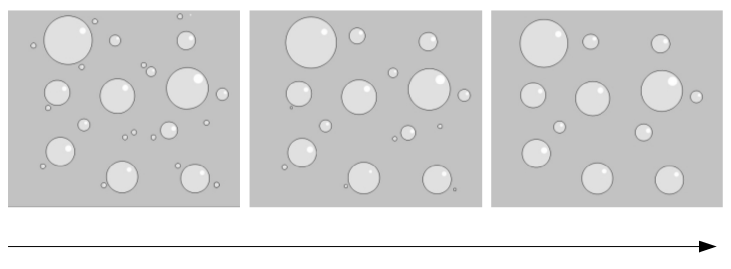}
\caption[]{Time evolution of gas bubbles in a liquid fluid.}
\label{ostwald-time-evolution}
\end{figure}

Smaller bubbles might {\it feed} larger bubbles -- the phenomenon of Ostwald ripening is ought to gas transfer from smaller bubbles to larger bubbles by diffusion in the liquid medium, provoking the radii increasing of larger bubbles at the expenses of decreasing radii of the smaller ones. (FIG.\ref{ostwald-time-evolution}).

The behaviour -- of a single spherical gas bubble at rest in a liquid fluid -- can be partially described by the Young-Laplace equation \cite{deGennes}: 
\be
\D P = P_{\rm in} - P_{\rm out} = \frac{2\g}{r}~, \label{young-laplace}
\ee
where $r$ is the bubble radius, $\g$ the surface tension and, $P_{\rm in}$ and $P_{\rm out}$ are the pressure inside (gas) and outside (liquid) the bubble, respectively. It shall be stressed that, from the Young-Laplace equation (\ref{young-laplace}), the gas bubble inner pressure ($P_{\rm in}$) is always greater than the outer pressure ($P_{\rm out}$), moreover, the smaller the bubble radius ($r$) the greater the pressure inside ($P_{\rm in}$) the bubble for a fixed ambient (outside) pressure ($P_{\rm out}$). 

\begin{figure}[h!]
\centering
\setlength{\unitlength}{1,0mm}
\includegraphics[width=5.2cm,height=2.65cm]{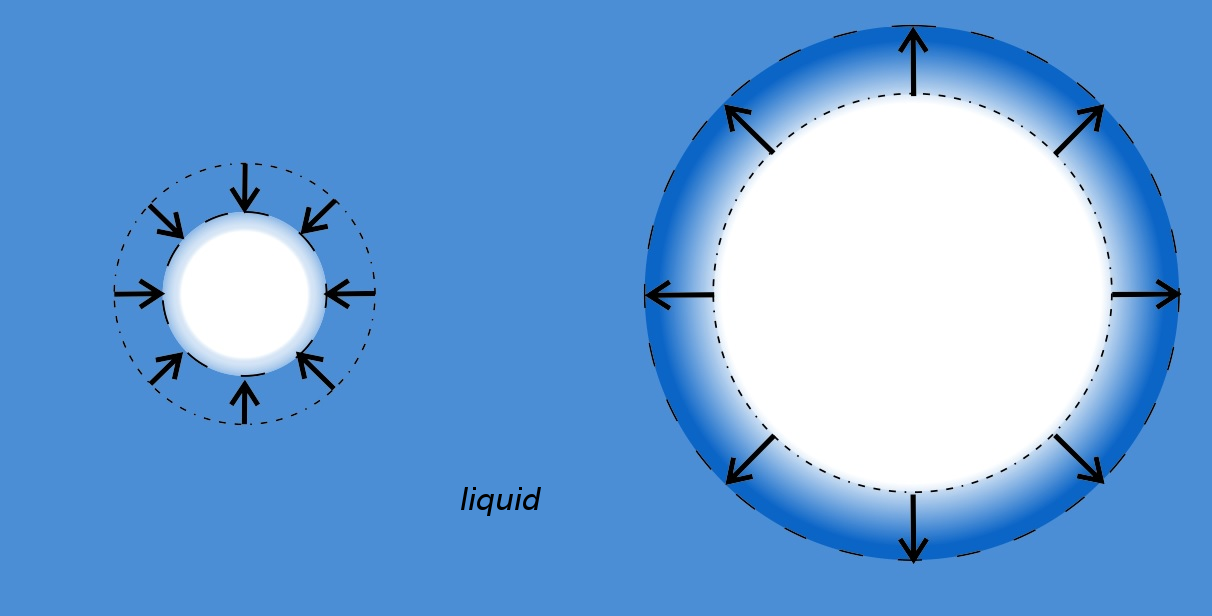}
\caption[]{Two gas bubbles with different radii in a liquid fluid.}
\label{two-bubbles}
\end{figure}

Now, what could happen if there were two gas bubbles, with different radii, into the liquid (FIG.\ref{two-bubbles})?  

Young and Laplace get the answer! Bearing in mind the Young-Laplace equation (\ref{young-laplace}) for the two bubbles, it follows that:  
\be
P_{r} = P_{\rm amb} + \frac{2\g}{r} \aand  P_{R} = P_{\rm amb} + \frac{2\g}{R}~, 
\label{young-laplace-two}
\ee
where $r$ and $R$ are de radii of the smaller and greater bubbles ($r<R$), respectively, the ambient (liquid) pressure is $P_{\rm amb}$, whereas $P_{r}$ is the inner pressure of the smaller bubble, and $P_{R}$ the inner pressure of the greater one. Therefore, it can be concluded from (\ref{young-laplace-two}) that, since $r<R$, then $P_{r}>P_{R}$, consequently, owing to the gradient pressure between the two bubbles, and by assuming a fixed ambient pressure (diver at a fixed depth -- $P_{\rm amb}=constant$), it stems -- gas flow from the smaller to the larger bubble -- the Ostwald ripening.

Dispersed throughout the body (inside tissues and blood) of a diver, astronaut, aviator or compressed air worker, there is a huge amount of gas (N$_2$, He, O$_2$, CO$_2$, H$_2$O vapour) bubbles whose radii vary from $10^{-1}\m$m to $10^{2}\m$m. The evolution of these gas bubbles is quite complicated since it involves altogether, compression/decompression, diffusion, isobaric counterdiffusion, coalescence, Ostwald ripening and  besides other more complex phenomena, therefore, experimental, theoretical and computational attempts to investigate, understand and describe such kind of complex system are herculean tasks. 

\section{Ostwald ripening: the experiment}
\label{experiment}
One of the main purpose of this work is to investigate and describe experimentally\footnote{The experiments were held at the Laboratory of Microfluidics and Complex Fluids of the Department of Physics.} the Ostwald ripening for gas bubbles \cite{bubble-group-article} -- the phenomenon of gas diffusion among bubbles -- in a liquid with some rheological parameters (density, surface tension and viscosity) as close as to the human blood \cite{blood}:
\ba
& {\rm density} \longrightarrow 1,00 \leq \r_{\rm blood} \leq 1,15~({\rm g}\,{\rm cm}^{-3})~,\nonumber\\
& {\rm surface~tension} \longrightarrow 15 \leq \g_{\rm blood} \leq 80~({\rm mN}\,{\rm m}^{-1})~,\nonumber\\
& {\rm viscosity} \longrightarrow 1,00 \leq \h_{\rm blood} \leq 4,00~({\rm mPa}\,{\rm s})~.\label{blood}
\ea 

\begin{figure}[h!]
\centering
\setlength{\unitlength}{1,0mm}
\includegraphics[width=5.2cm,height=2.65cm]{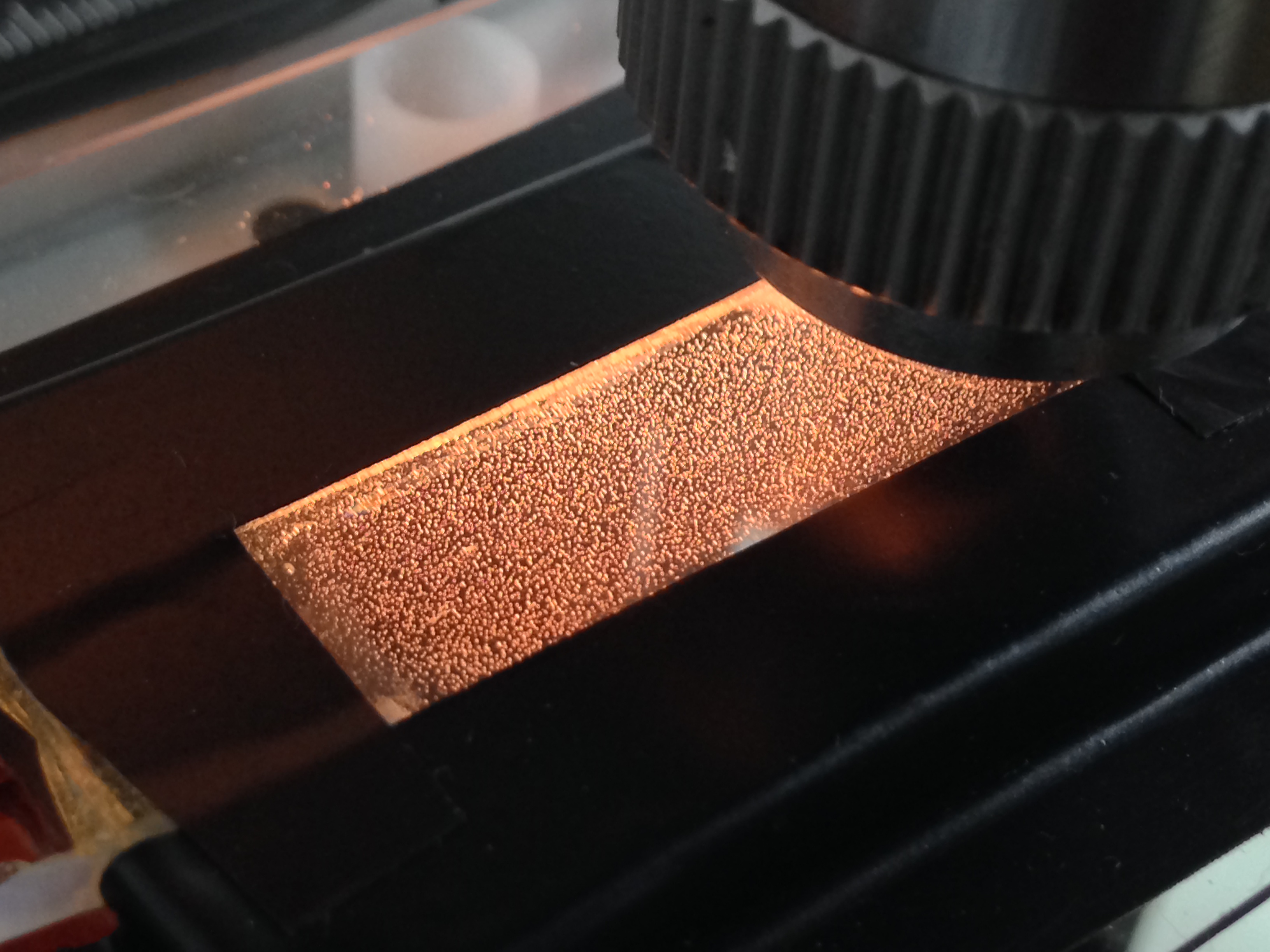}
\caption[]{Air bubbles confined in a chamber.}
\label{bubble-chamber}
\end{figure}

The physical quantities adopted, in order to describe the time evolution of the whole system consisting of air bubbles in a liquid solution (v/v) -- 75\% glycerol + 25\% H$_2$O (deionized) -- confined in a (bubbles) chamber (FIG.\ref{bubble-chamber}), are: mean bubble radius ($\ov{R}(t)$), number of bubbles ($N(t)$), radii (frequency) distribution ($f(R,t)$) and radii normalized (probability) distribution ($p(R,t)$).

\begin{figure}[t!]
\centering
\setlength{\unitlength}{1,0mm}
\includegraphics[width=5.2cm,height=5.6cm]{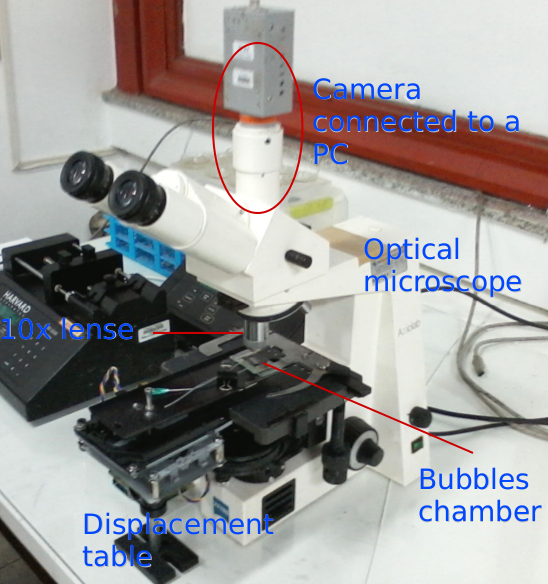}
\caption[]{Experimental apparatus.}
\label{apparatus}
\end{figure}

The experiment apparatus includes an optical microscope (10x lense), a B\&W camera coupled to the microscope and connected to a computer, a bubbles chamber attached to a displacement table controlled by the computer (FIG.\ref{apparatus}), also, there is a mixer used to produce (by cavitation) the air bubbles in the liquid solution, prior to injection into the bubbles chamber (FIG.\ref{bubble-chamber}). 

\begin{figure}[h!]
\centering
\setlength{\unitlength}{1,0mm}
\includegraphics[width=5.2cm,height=4.2cm]{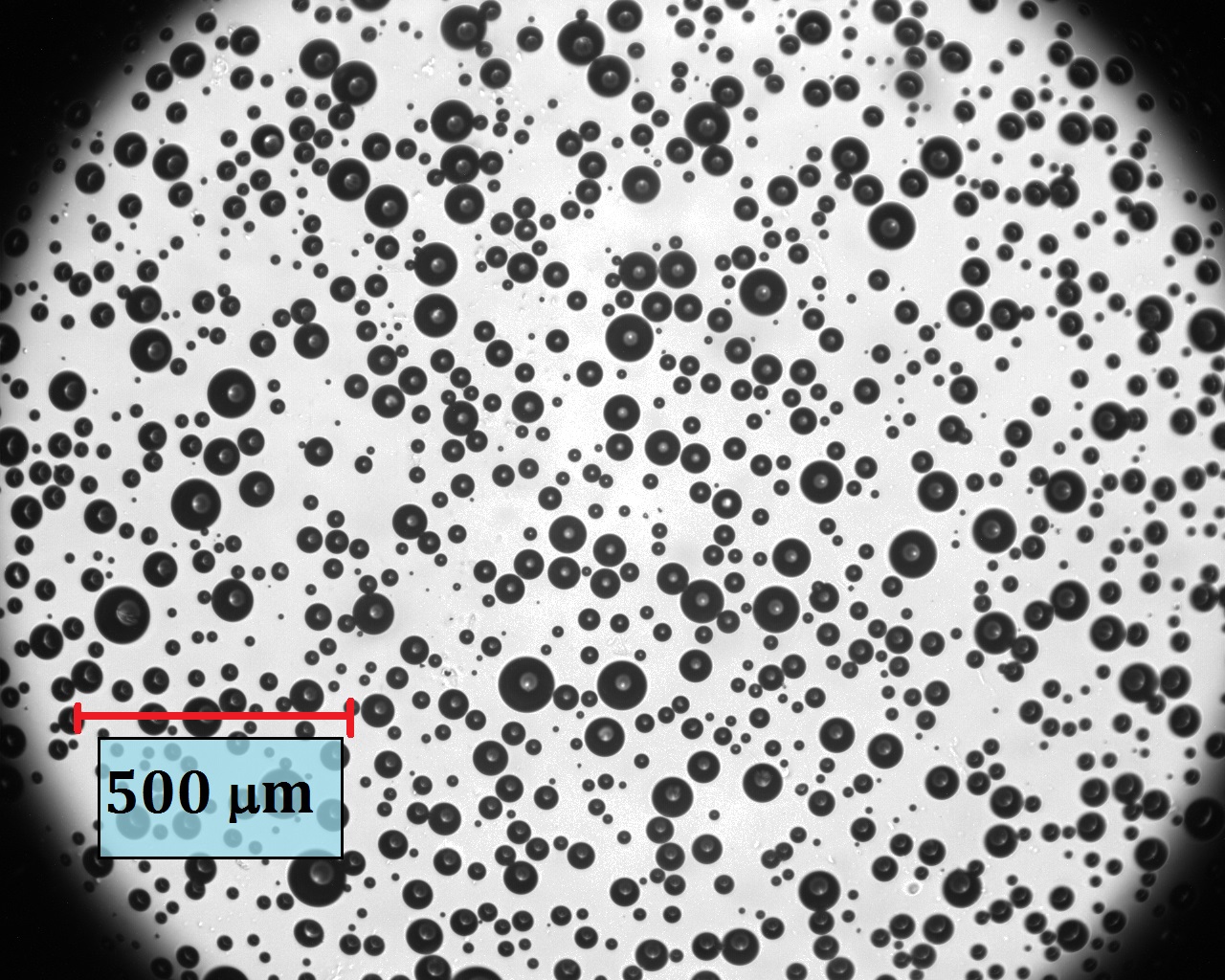}
\caption[]{Air bubbles: microscope focus.}
\label{microscope-focus}
\end{figure}

The liquid solution (v/v) used in the experiment contains 75\% of glycerol and 25\% of deionized water, which exhibits the following measured rheological parameters:
\ba
& {\rm density} \longrightarrow \r_{\rm exp} = (1,17 \pm 0,01)~{\rm g}\,{\rm cm}^{-3}~,\nonumber\\
& {\rm surface~tension} \longrightarrow \g_{\rm exp} = (65,3 \pm 0,01)~{\rm mN}\,{\rm m}^{-1}~,\nonumber\\
& {\rm viscosity} \longrightarrow \h_{\rm exp} = (34,530 \pm 0,002)~{\rm mPa}\,{\rm s}~,\label{exp}
\ea 
at $25^{\circ}{\rm C}$ room temperature. It shall be noticed that, in order to set apart (experimentally) the Ostwald ripening from other effects, namely, by reducing potential coalescence among air bubbles and also by avoiding their dislocations (FIG.\ref{microscope-focus}), the liquid solution viscosity, $\h_{\rm exp}$ (\ref{exp}), had to be fixed one order of magnitude greater than the mean blood viscosity, $\h_{\rm blood}$ (\ref{blood}).

\begin{figure}[h!]
\centering
\setlength{\unitlength}{1,0mm}
\includegraphics[width=9.0cm,height=6.0cm]{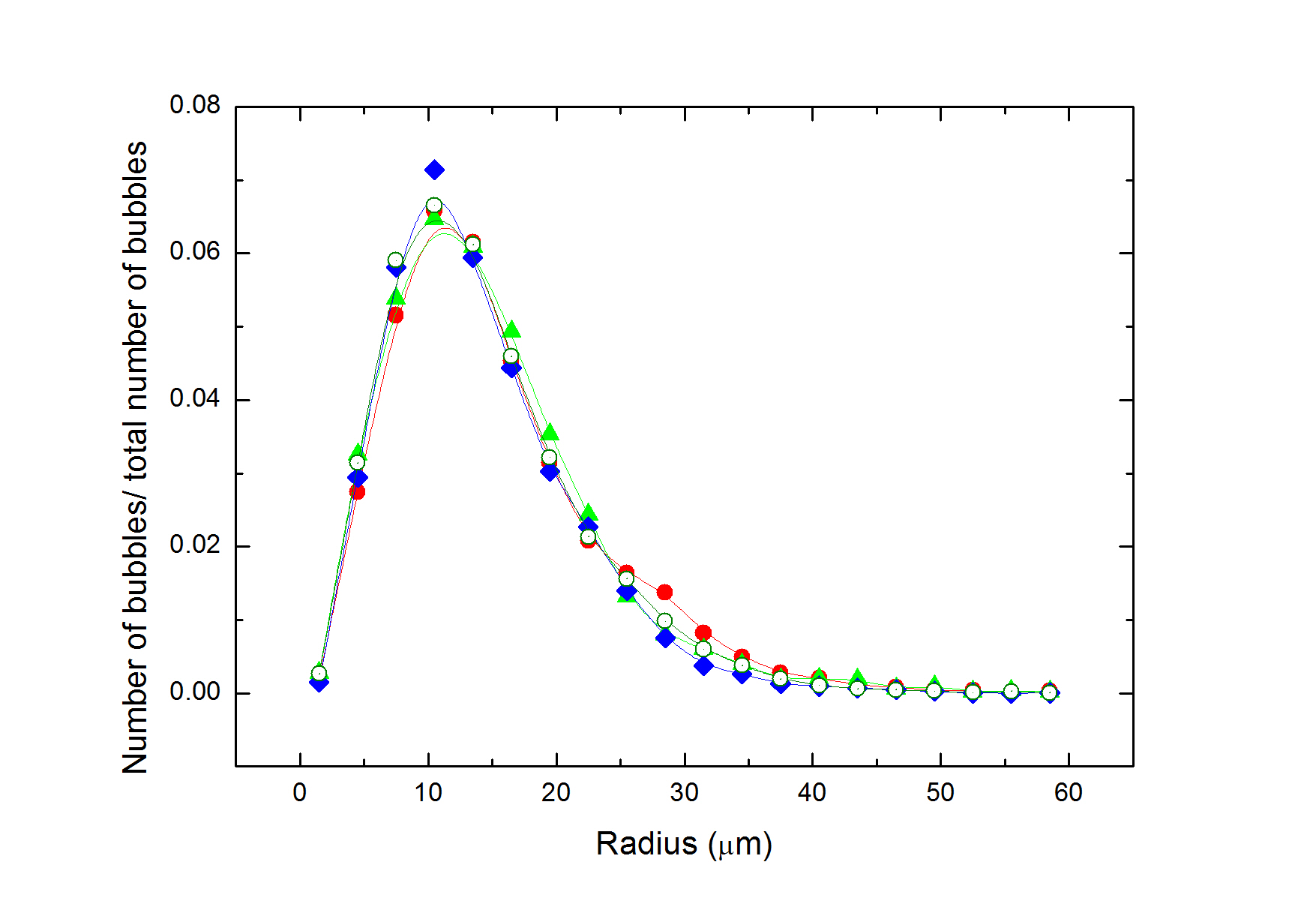}
(a)\\
\includegraphics[width=9.0cm,height=6.0cm]{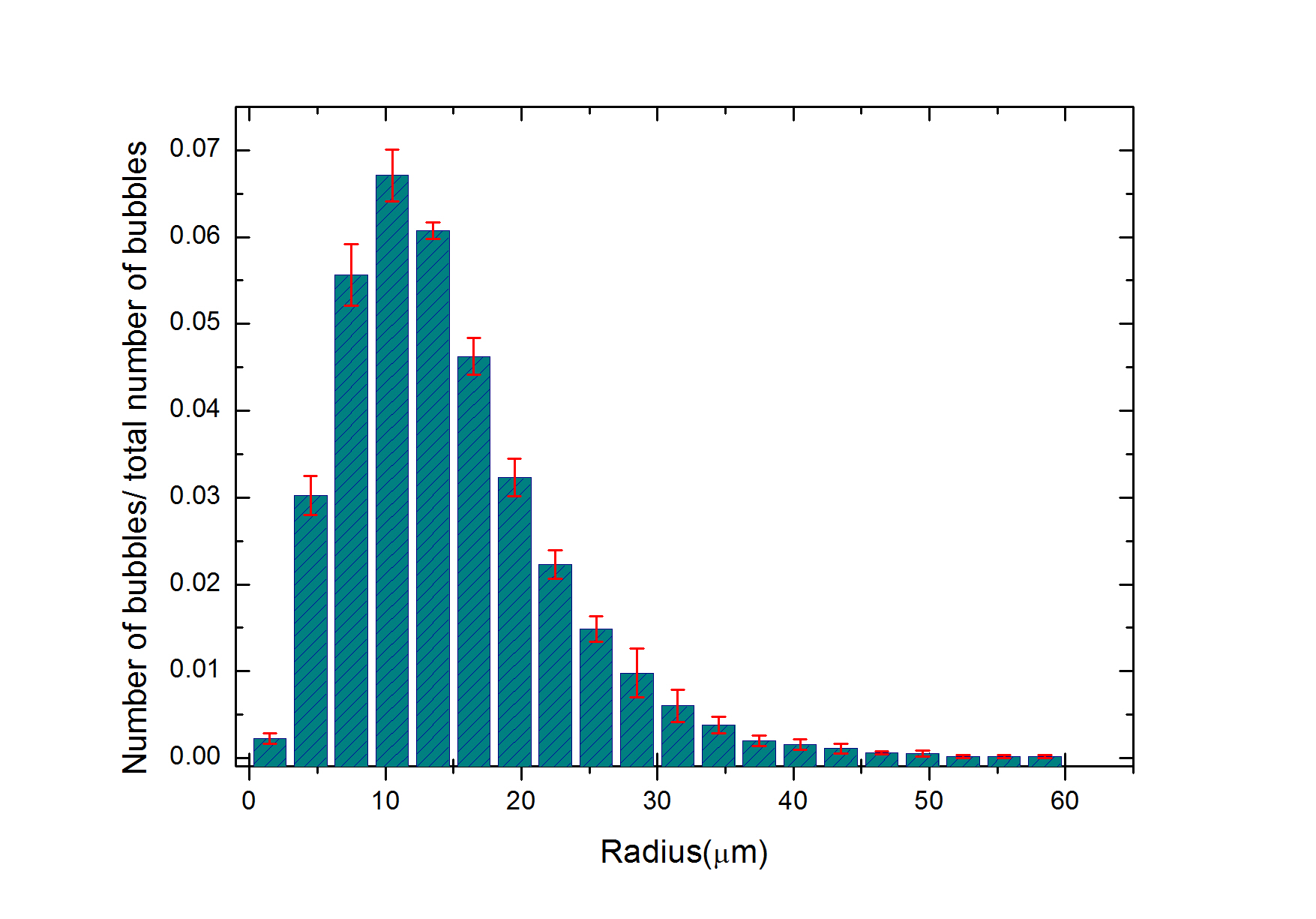}
(b)
\caption[]{Reproducibility: (a) the initial radii normalized distribution ($p(R,0)$) for the four samples; (b) the respective histogram with error bars.}
\label{four-initial-distributions}
\end{figure}

The experiment runs for four samples of air bubbles in the liquid solution, 75\% of glycerol and 25\% of deionized water (v/v). The initial radii normalized distribution ($p(R,0)$) -- defined by the ratio among the number of bubbles with radius $R$ (the radii distribution $f(R,0)$) and the total number of bubbles ($N(0)\sim 10^4$) at instant zero ($0$h) -- of the four samples analyzed, validates the reproducibility of the experiment (FIG.\ref{four-initial-distributions}). 

\begin{figure*}[t]
\centering
\setlength{\unitlength}{1,0mm}
\includegraphics[width=9.0cm,height=6.0cm]{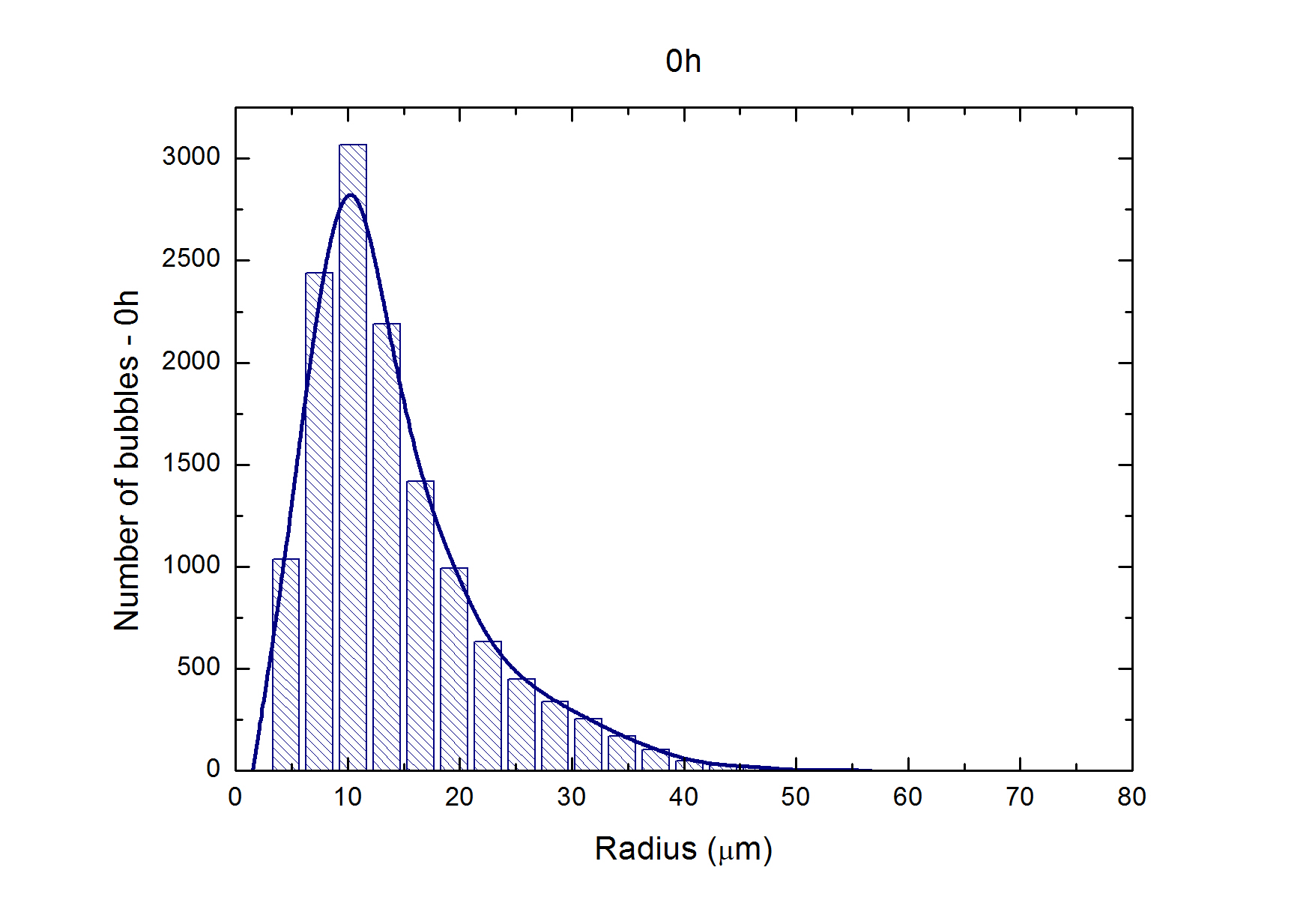}\includegraphics[width=9.0cm,height=6.0cm]{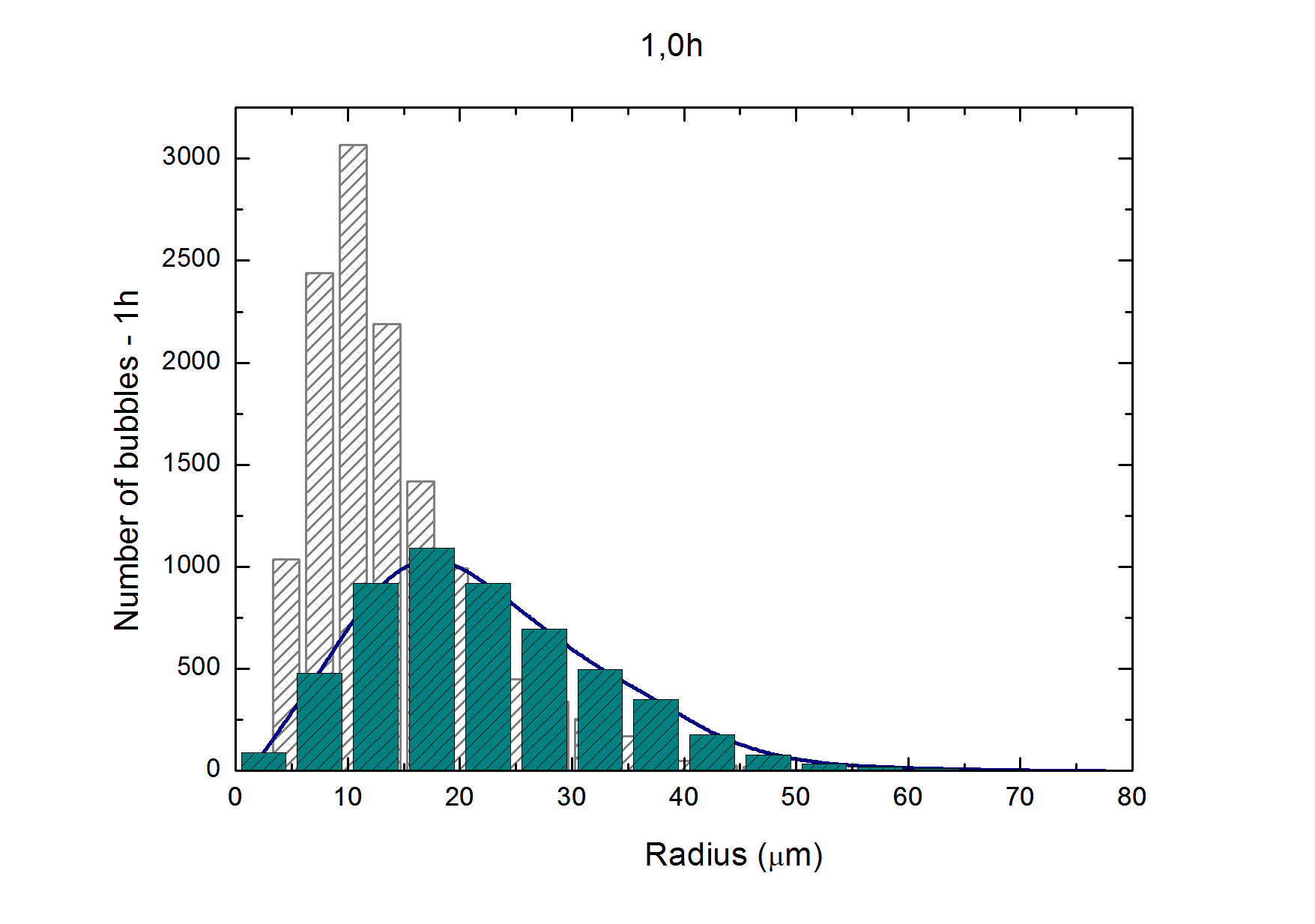}
\includegraphics[width=9.0cm,height=6.0cm]{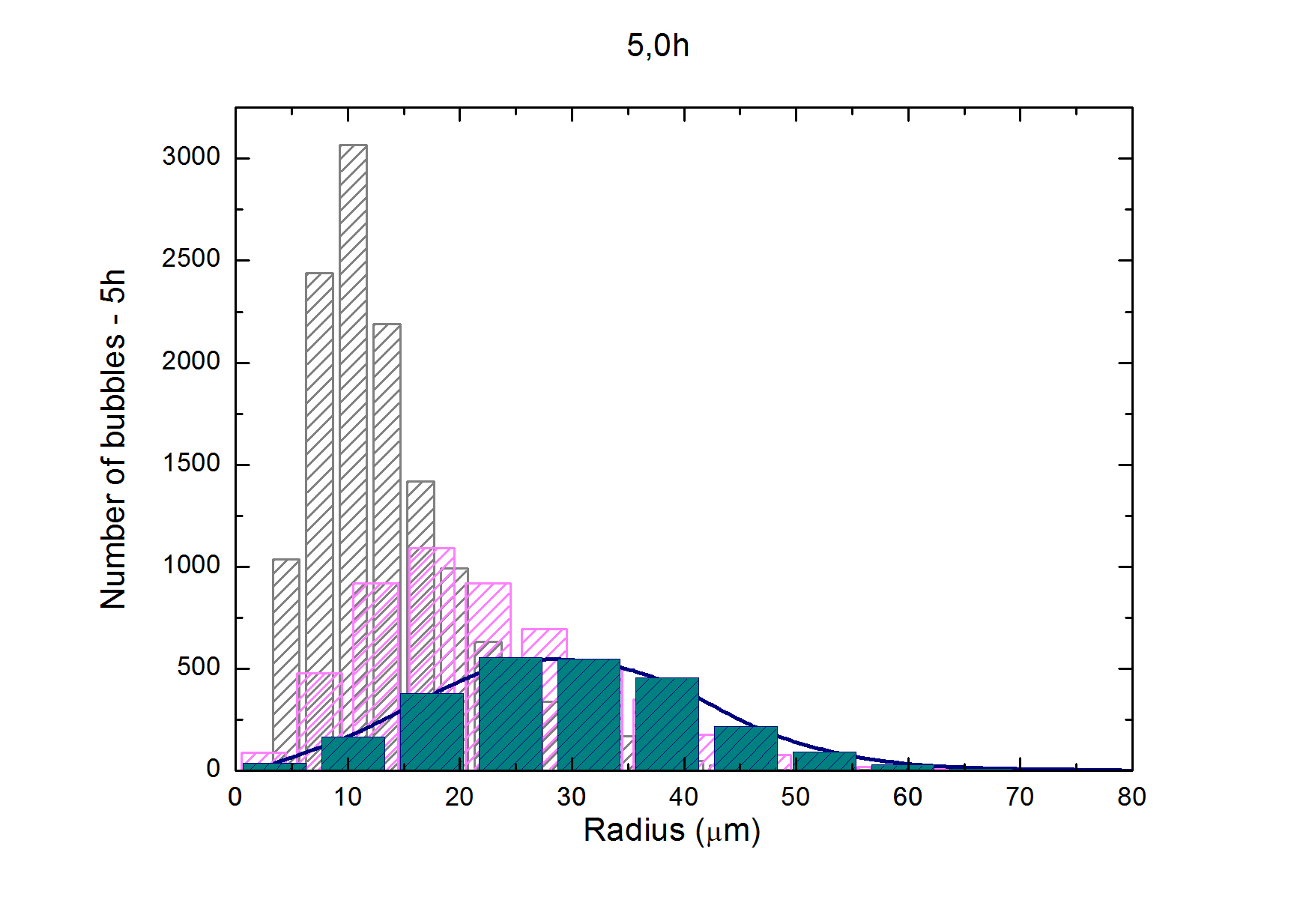}\includegraphics[width=9.0cm,height=6.0cm]{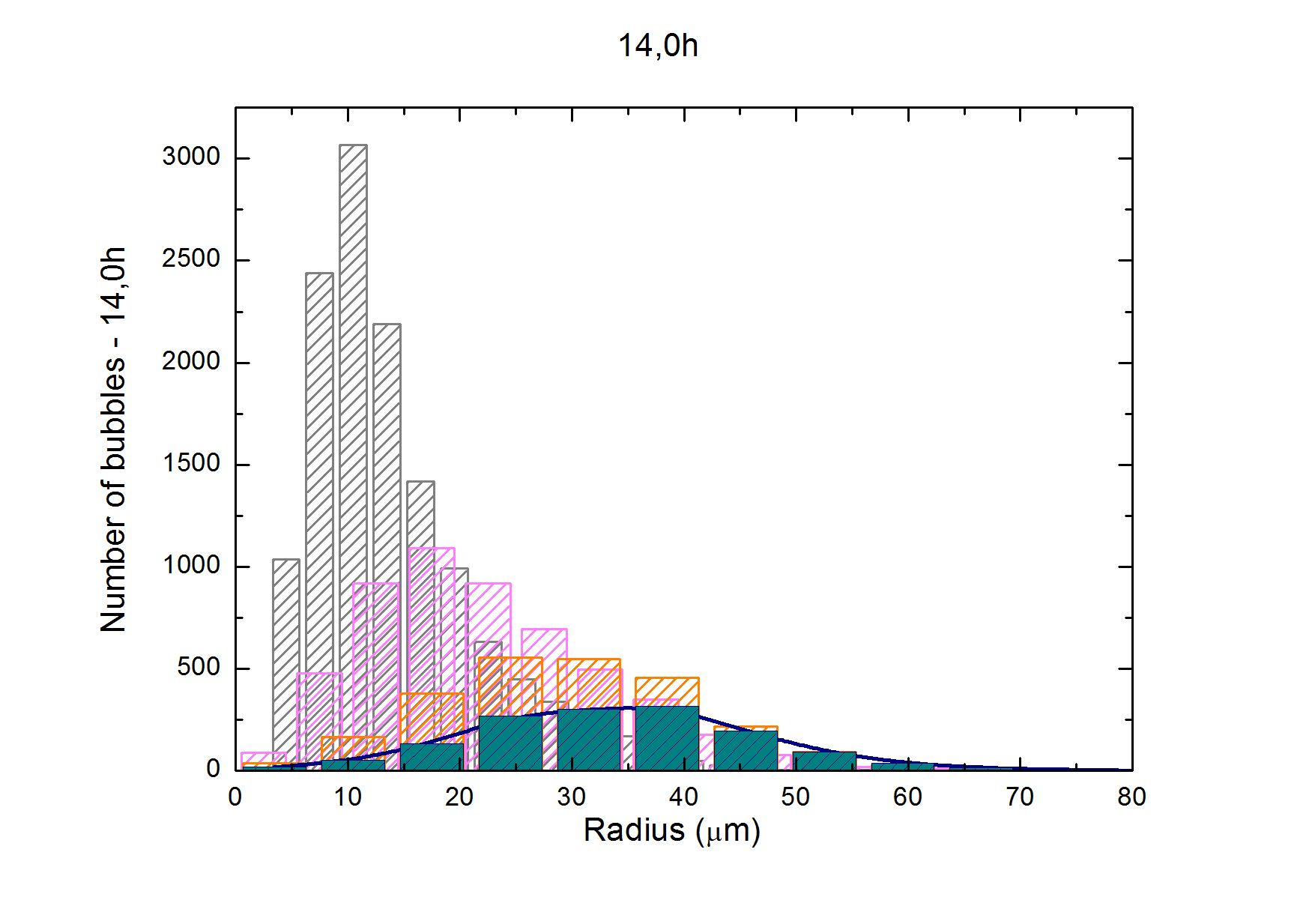}
\caption[]{The radii (frequency) distribution ($f(R,t)$) at $t=0$h, $t=1$h, $t=5$h and $t=14$h.}
\label{distributions}
\end{figure*}

\begin{figure*}[t]
\centering
\setlength{\unitlength}{1,0mm}
\includegraphics[width=9.0cm,height=6.0cm]{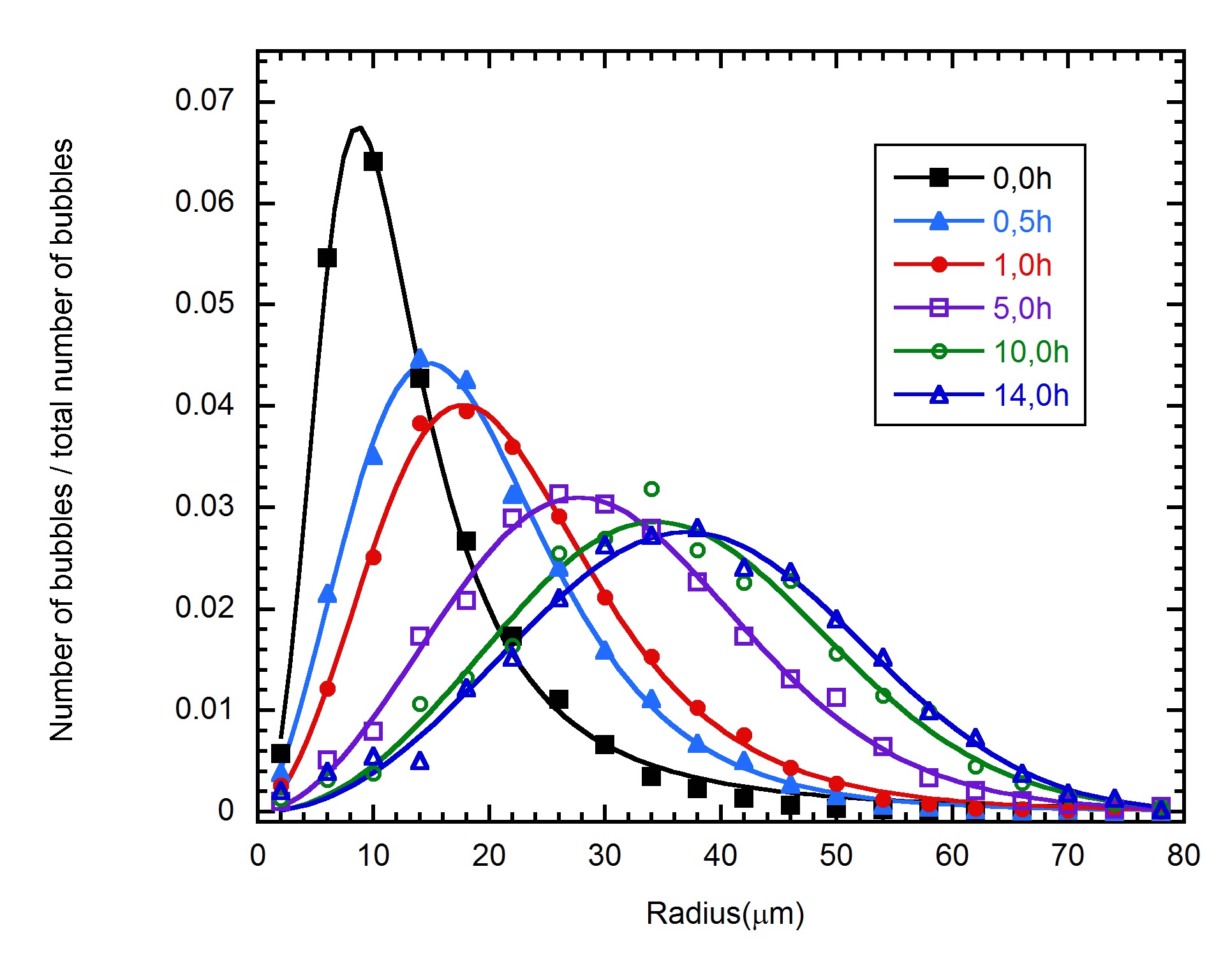}
\caption[]{The radii normalized (probalility) distribution ($p(R,t)$) at $t=0$h, $t=1$h, $t=5$h and $t=14$h.}
\label{normalized-distribution}
\end{figure*}

The radii (frequency) distribution as a function of time ($f(R,t)$) is acquired analyzing\footnote{The software used to treat the pictures, so as to measure the radii of the bubbles, was the open source image processing  {\bf ImageJ} \cite{imagej}.} the images taken by the B\&W camera (coupled to the microscope) at different zones of the bubbles chamber -- which are reached through the displacement table controlled by the computer. When the acquisition of the images by the B\&W camera and the subsequent bubbles radii measurement, the radii (frequency) distribution ($f(R,t)$) and the radii normalized (probability) distribution ($p(R,t)$) were obtained at the time instants: $0$h, $1$h, $5$h and $14$h (FIG.\ref{distributions} and FIG.\ref{normalized-distribution}). In what concerns the bubbles time evolution, it can be deduced, from FIG.\ref{distributions} and FIG.\ref{normalized-distribution}, that the number of bubbles ($N(t)$) decreases whereas mean bubble radius ($\ov{R}(t)$) increases, which shall be proved below. 

\begin{figure}[t]
\centering
\setlength{\unitlength}{1,0mm}
\includegraphics[width=9.0cm,height=6.0cm]{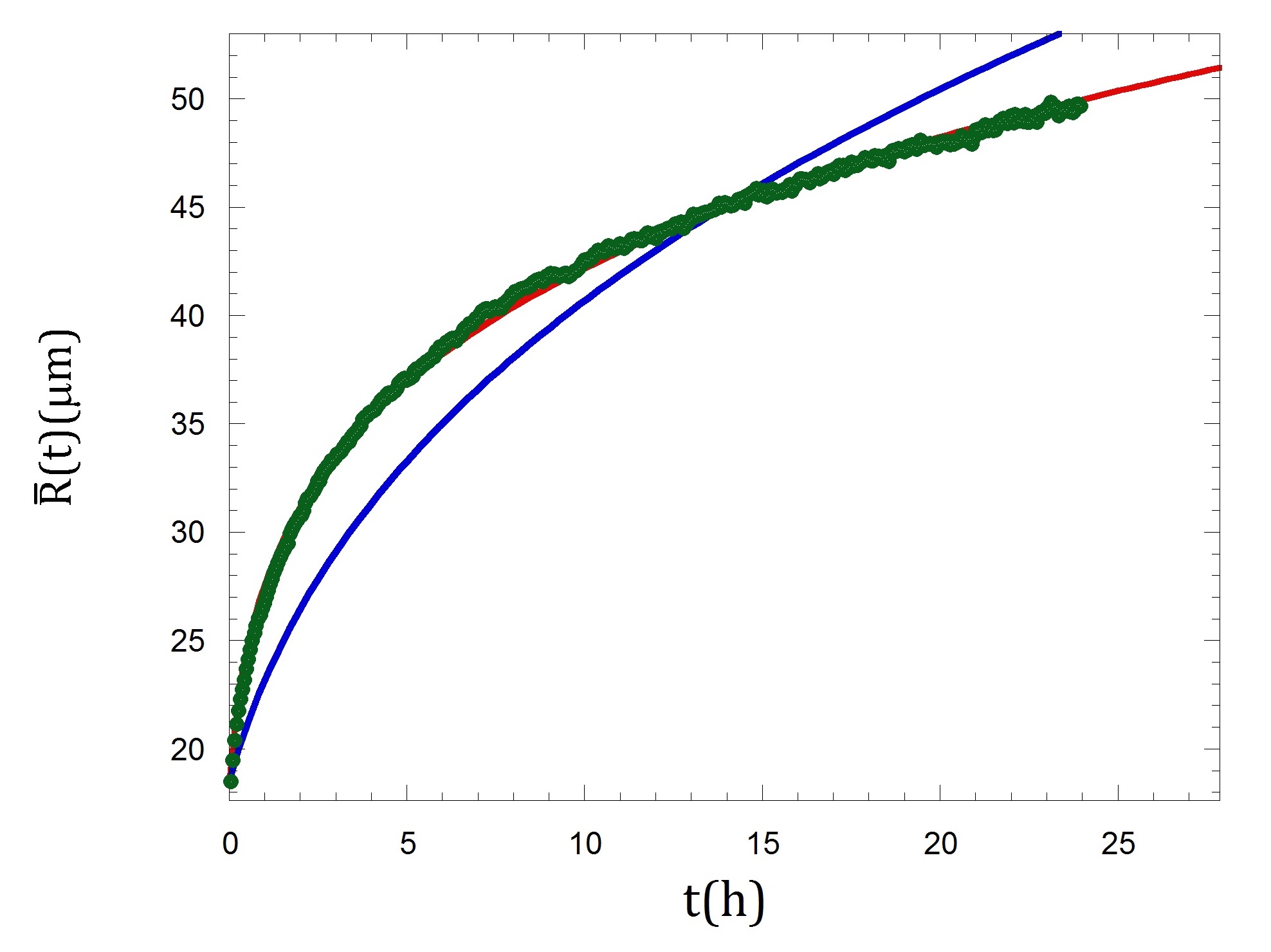}
\caption[]{The mean bubble radius: the experimental data fit ($\ov{R}(t)$ -- red line) and the LSW theory prediction ($\ov{R}_{\rm LSW}(t)$ -- blue line) assuming the same experimental initial condition, the initial mean bubble radius 
$\ov{R}(0)=18,42\m$m.}
\label{mean-radius}
\end{figure}

The experimental results for the mean bubble radius ($\ov{R}(t)$) show that it increases monotonically in time, which can be straightforwardly concluded from FIG.\ref{mean-radius}, wherein, together with the experimental curve fit, it is also sketched a curve for the mean bubble radius ($\ov{R}_{\rm LSW}(t)$) if the bubbles dynamics was dictated by the LSW (Lifshitz-Slyozov-Wagner) theory \cite{lsw}. The mean bubble radius ($\ov{R}_{\rm LSW}(t)$) in LSW theory is given by:
\ba
&\ov{R}_{\rm LSW}(t) = \left[\ov{R}^3(0) + Kt\right]^{\frac{1}{3}} ~,~~ \ov{R}(0)=18,42\m{\rm m} \nonumber\\ 
&\aand K=6,1\times 10^3{\rm m}^3{\rm s}^{-1} ~, \nonumber 
\ea
where $K$ depends on the temperature, surface tension, diffusion coefficient, gas solubility and gas molar volume. Therefore, it can be verified from FIG.\ref{mean-radius} (blue line) that the LSW theory does not properly describe such a system of gas bubbles in a liquid, which should be expected since the system analyzed does not satisfy the primary premises assumed by the LSW theory. Nevertheless, bearing in mind the experimental data acquired and modelling the mean bubble radius ($\ov{R}(t)$) as below:
\ba
&\ov{R}(t) = \left[\ov{R}^{\frac{1}{\chi}}(0) + Kt\right]^{\chi} ~, \label{rbar-time}\\
&\ov{R}(0)=18,42\m{\rm m} ~,~~ K=2,0\times 10^7{\rm m}^{\frac{1}{\chi}}{\rm s}^{-1} \nonumber\\ 
&\aand \chi=0,1956 ~, \nonumber 
\ea
from FIG.\ref{mean-radius} (red line) it follows that the empirical model (\ref{rbar-time}) proposed perfectly fits the experimental data.  

\begin{figure}[t]
\centering
\setlength{\unitlength}{1,0mm}
\includegraphics[width=9.0cm,height=6.0cm]{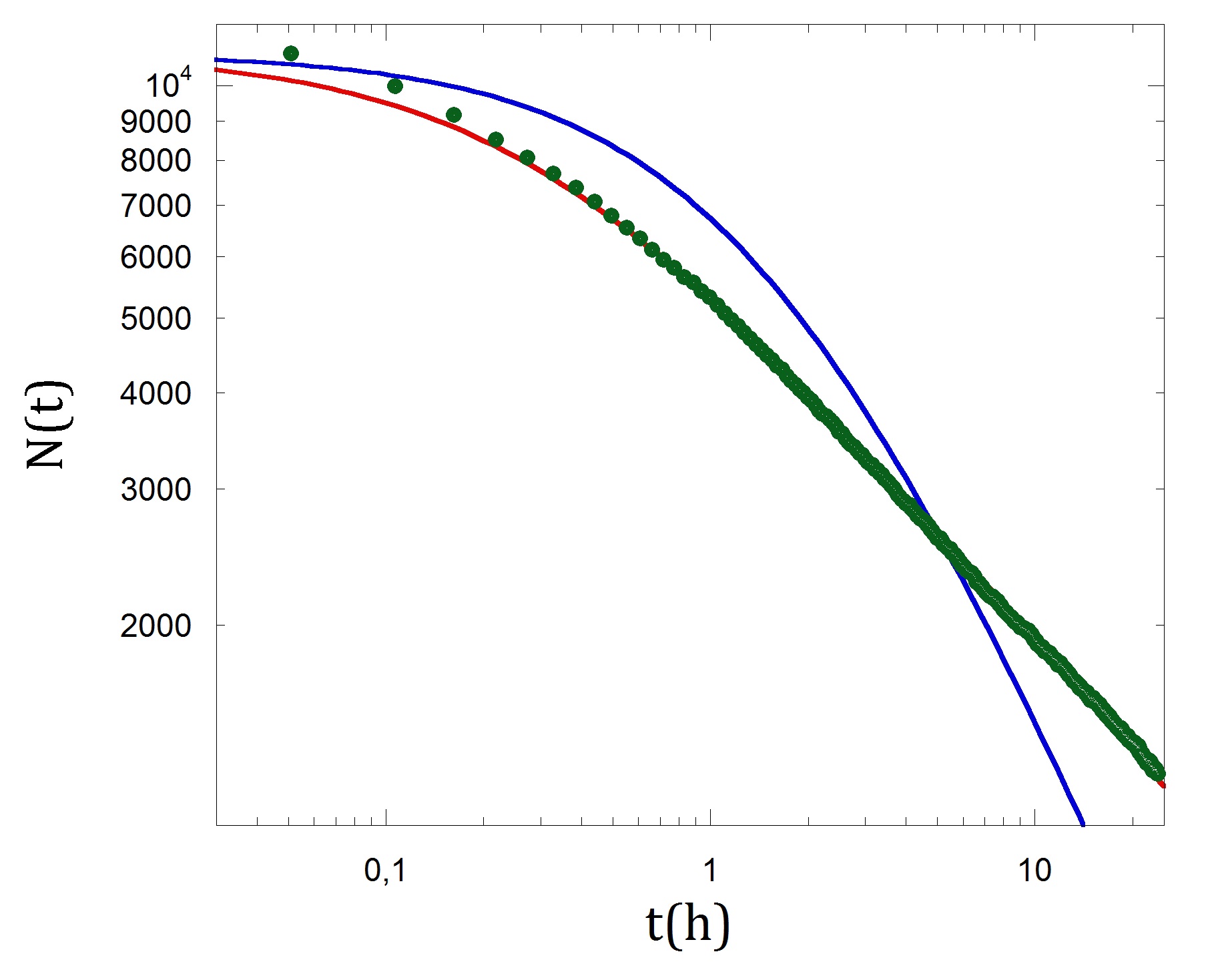}
\caption[]{The number of bubbles: the experimental data fit ($N(t)$ -- red line) and the LSW theory prediction ($N_{\rm LSW}(t)$ -- blue line) assuming the same experimental initial condition, the mean bubble radius, $\ov{R}(0)=18,42\m$m, and the number of bubbles, $N(0)=1,8\times 10^4$.}
\label{number}
\end{figure}

On the other way around, the experiment shows that, while the mean bubble radius ($\ov{R}(t)$) increases in time, the number of bubbles ($N(t)$) decreases monotonically as displayed in FIG.\ref{number}, where it can also be seen that the number of bubbles ($N_{\rm LSW}(t)$) proposed by the LSW theory: 
\ba
&N_{\rm LSW}(t) = N(0) \displaystyle\frac{\ov{R}^3(0)}{\ov{R}^3(0) + Kt} ~,~~ \ov{R}(0)=18,42\m{\rm m}~, \nonumber\\ 
&N(0)=1,8\times 10^4 \aand K=6,1\times 10^3{\rm m}^3{\rm s}^{-1} ~, \nonumber 
\ea
does not fit (blue line) the experimental data. However, by considering the experimental data, and modelling the number of bubbles ($N(t)$) as follows:
\ba
&{N}(t) = N(0) \displaystyle\frac{\ov{R}^{\frac{\l}{\chi}}(0)}{\left[\ov{R}^{\frac{1}{\chi}}(0) + Kt\right]^{\l}} ~, \label{n-time} \\ 
&\ov{R}(0)=18,42\m{\rm m} ~,~~ K=2,0\times 10^7{\rm m}^{\frac{1}{\chi}}{\rm s}^{-1} ~,\nonumber\\
&\chi=0,1956 \aand \l=0,48 ~, \nonumber 
\ea
it can be verified that the empirical model (\ref{n-time}) fits the experimental data, FIG.\ref{number} (red line). 

In summary, the experiment realized -- upon a system of air bubbles in a liquid fluid with some rheological parameters close to the human blood -- shows that the mean bubble radius ($\ov{R}(t)$) increases (FIG.\ref{mean-radius}) in time whereas the number of bubbles ($N(t)$) decreases (FIG.\ref{number}). Moreover, 
it is verified straightforwardly that the smaller bubbles disappear whereas the larger bubbles, the ones potentially dangerous for the diver which cause the decompression sickness, grow up. 

Based on the experimental analysis of the Ostwald ripening phenomenon, it has been proposed an empirical model so as to describe the time evolution of the mean bubble radius (\ref{rbar-time}) and the number of bubbles (\ref{n-time}), where it properly describes the time evolution of the system. 

Taking into consideration the Ostwald ripening empirical model proposed here, it should be interesting to probe its implementation to the Reduced Gradient Bubble Model (RGBM) in such a manner to compute decompression sickness risks and develop dive tables for further diving tests. 

In addition to the experimental results and the empirical model introduced here, simulation of gas bubbles evolution has been performed and some preliminary results are presented in the next Section. 

\section{Ostwald ripening: the simulation}
\label{simulation}

Another purpose of this work is, by adopting the finite element method, the computational modelling and simulation\footnote{The computational simulations were held at the Complex Systems Investigation Laboratory  of the Department of Physics.} of gas diffusion among bubbles in a liquid and their time evolution. By working out the Ostwald ripening computational simulation together with the experimental results it stems a more profound knowledge about the details of the Ostwald ripening phenomenon, which are crucial to its further implementation to the RGBM so as to realize diving field tests. 

At this preliminary stage, the simulations are performed by assuming nitrogen (N$_2$) gas bubbles into water, where its initial N$_2$ gas concentration is equivalent to critical radius given by $R_{\rm c}=10\m{\rm m}$, which means that a single N$_2$ bubble with a radius equal to the critical radius would remain in equilibrium, namely, it would neither decreases nor increases.

\begin{figure}[h!]
\centering
\setlength{\unitlength}{1,0mm}
\includegraphics[width=4.25cm,height=4.25cm]{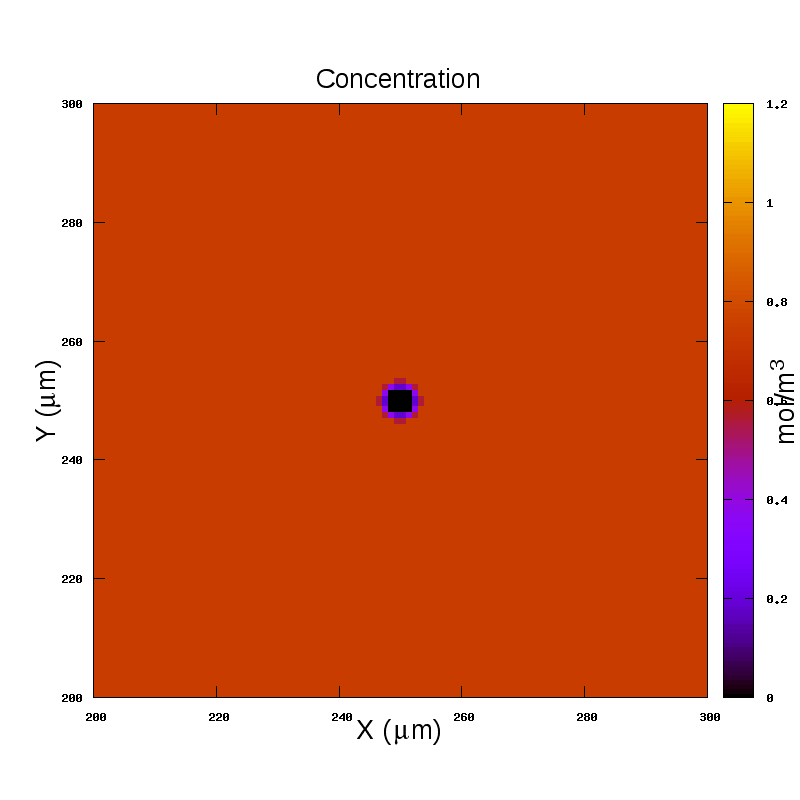}
\includegraphics[width=4.25cm,height=4.25cm]{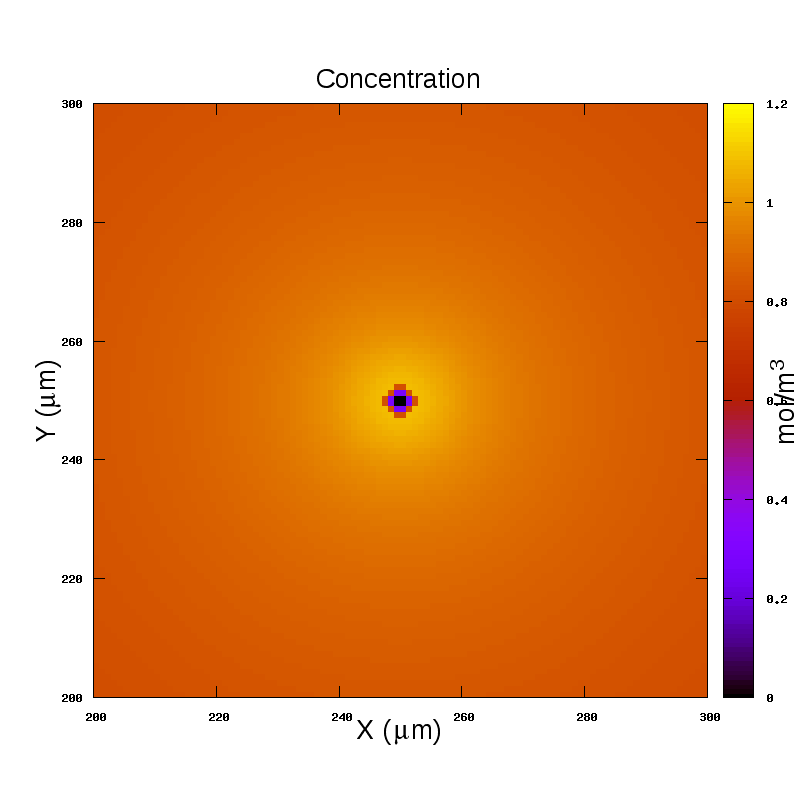}\\
\includegraphics[width=6.0cm,height=4.25cm]{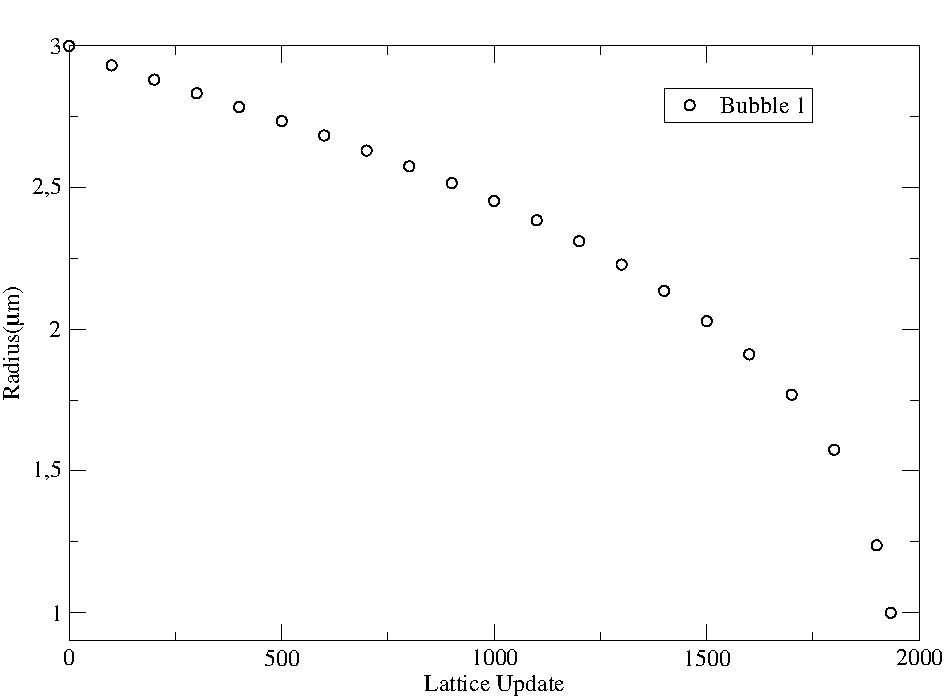}
\caption[]{Simulation of one bubble ($R_{\rm c}=10\m{\rm m}$).}
\label{one-bubble}
\end{figure}

The first case considered deals with one single N$_2$ bubble (FIG.\ref{one-bubble}), in a $500\m{\rm m}\times500\m{\rm m}$ lattice, with its initial radius given by $R(0)=3\m{\rm m}$. Therefore, as expected, it decreases in time, since its radius is smaller than the critical one, $R_{\rm c}=10\m{\rm m}$. 

\begin{figure}[h!]
\centering
\setlength{\unitlength}{1,0mm}
\includegraphics[width=4.25cm,height=4.25cm]{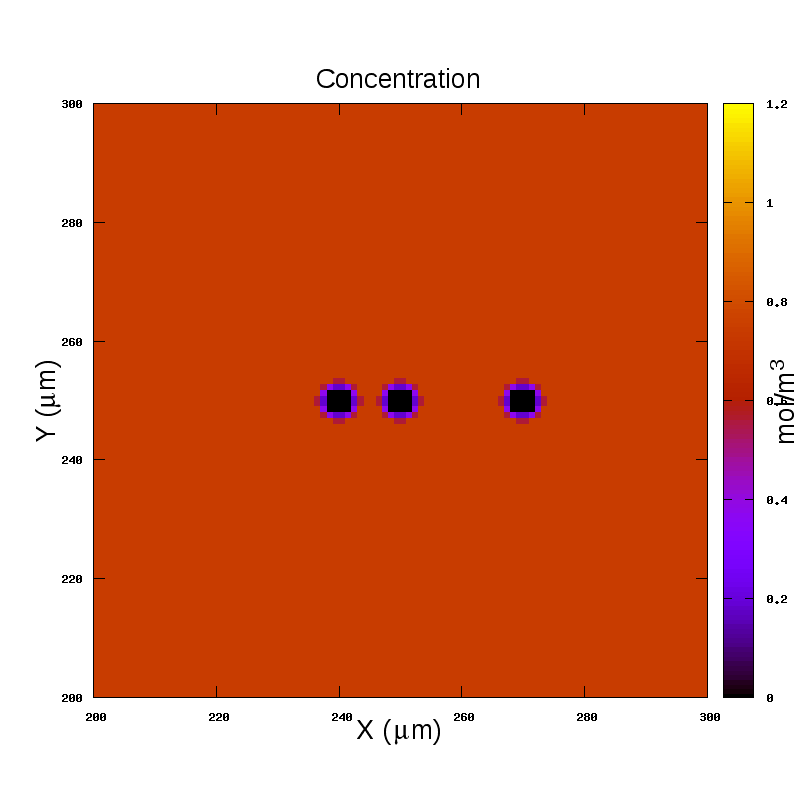}
\includegraphics[width=4.25cm,height=4.25cm]{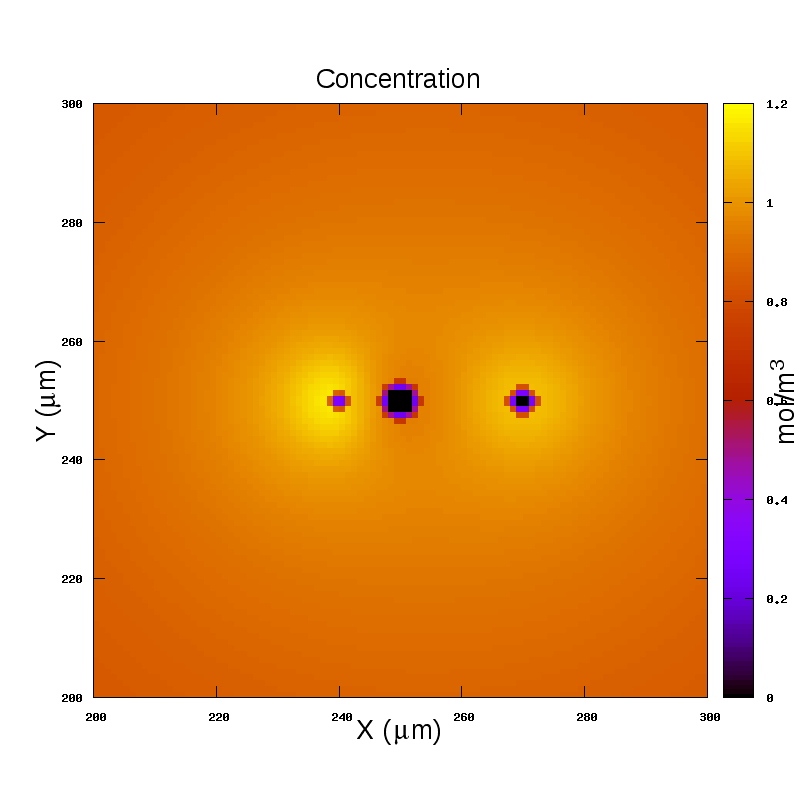}\\
\includegraphics[width=6.0cm,height=4.25cm]{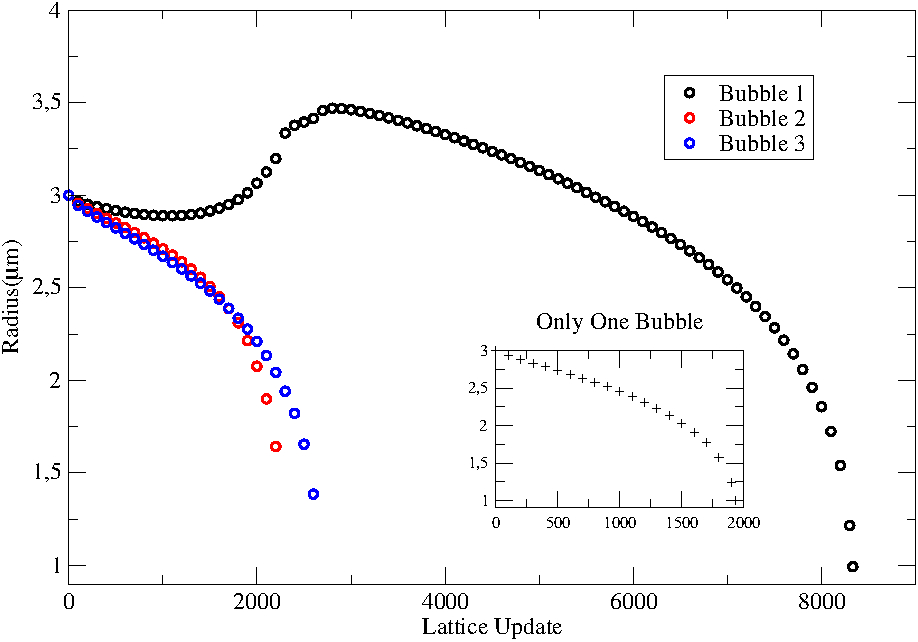}
\caption[]{Simulation of three bubbles ($R_{\rm c}=10\m{\rm m}$).}
\label{three-bubbles}
\end{figure}

A configuration of three bubbles aligned are the second case studied, where they have the same initial radius $R(0)=3\m{\rm m}$ but are spaced such that (FIG.\ref{three-bubbles}) the left bubble (bubble 2 -- red dots) are closer to the central bubble (bubble 1 -- black dots) than the right bubble (bubble 3 -- blue dots), with the lattice size equal to $500\m{\rm m}\times500\m{\rm m}$. There, it is verified that the closer bubble (bubble 2 -- red dots) to the central,  decreases faster than the farther bubble (bubble 3 -- blue dots), however, while they decrease the central bubble increases until they disappear, thereafter the central bubble starts decreasing. It shall be stressed that unlike the LSW theory -- where all three bubbles would decrease simultaneously, namely with the same time rate, independently on how they are spaced -- the behaviour of the three bubbles are strongly dictated by the distance among them, becoming evident that the inter-bubbles gas pressure gradient should be taken into consideration in a future extension of the Reduced Gradient Bubble Model (RGBM). 

\begin{figure}[h!]
\centering
\setlength{\unitlength}{1,0mm}
\includegraphics[width=4.25cm,height=4.25cm]{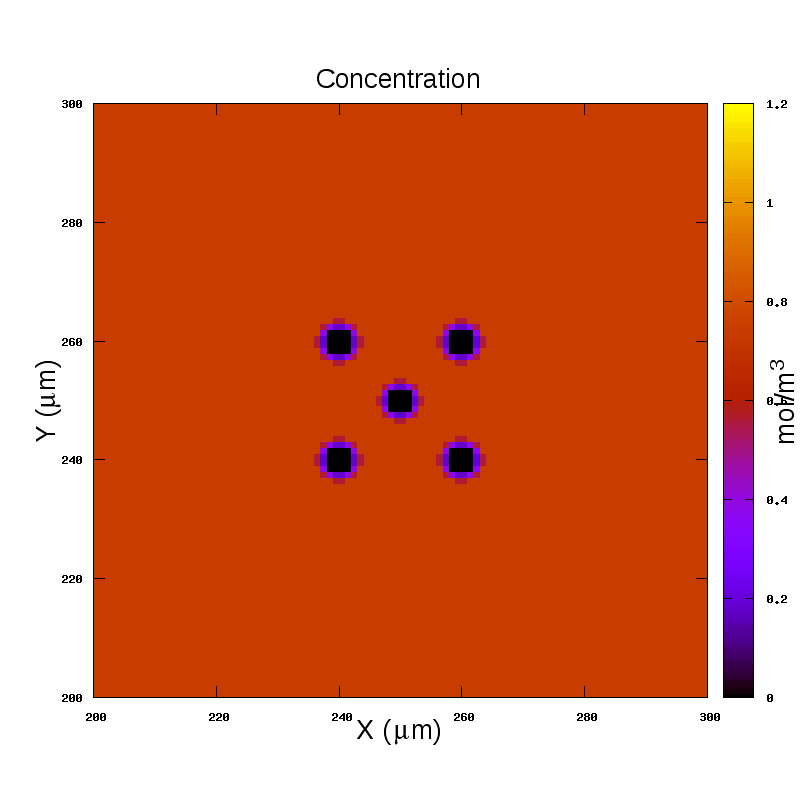}
\includegraphics[width=4.25cm,height=4.25cm]{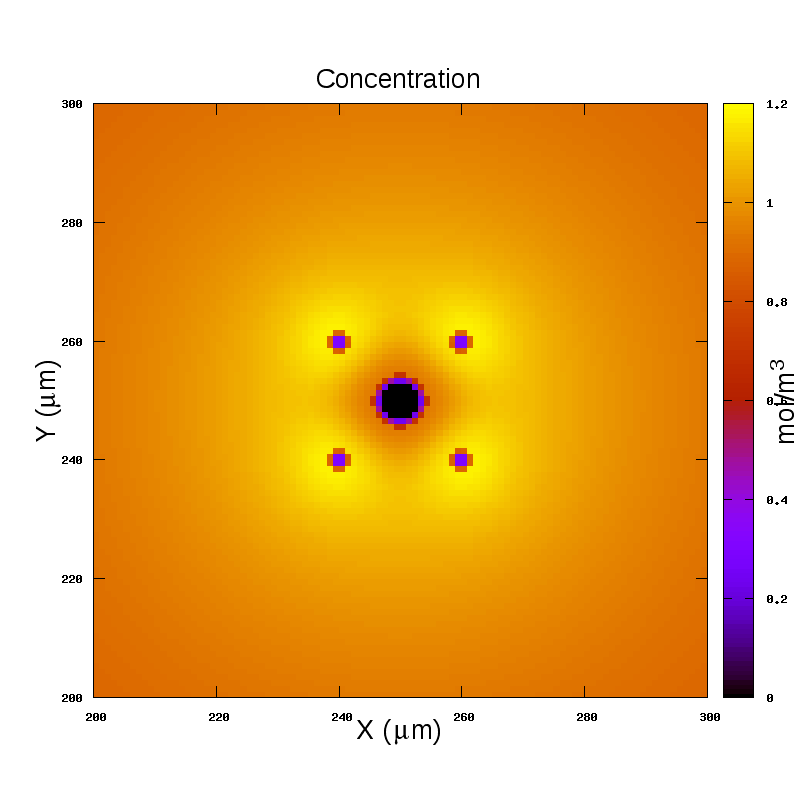}\\
\includegraphics[width=6.0cm,height=4.25cm]{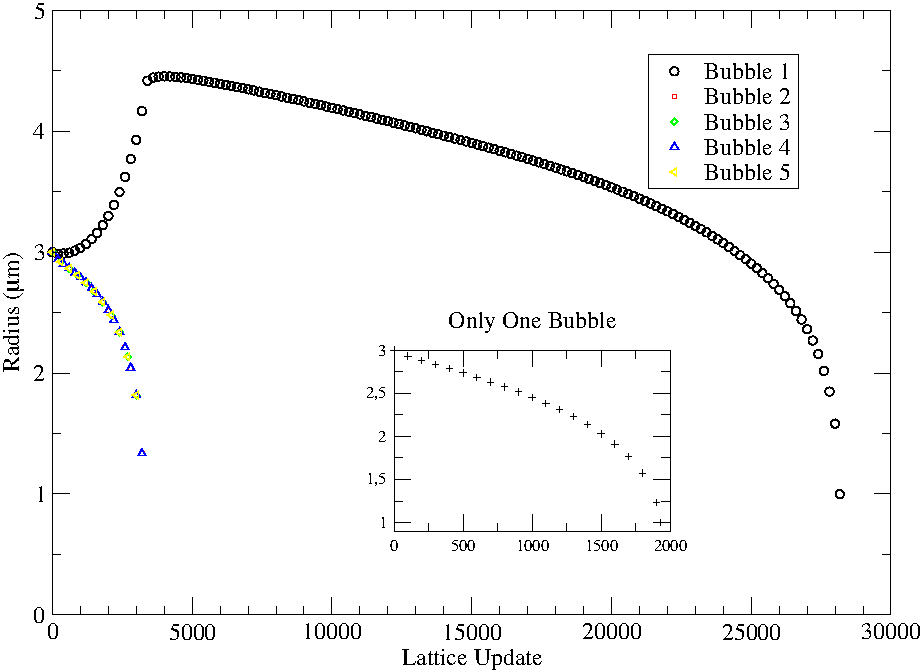}
\caption[]{Simulation of five bubbles ($R_{\rm c}=10\m{\rm m}$).}
\label{five-bubbles}
\end{figure}

The third case analyzed is a system of five bubbles, in a $500\m{\rm m}\times500\m{\rm m}$ lattice, all with the same initial radius $R(0)=3\m{\rm m}$, arranged so that four of the five bubbles are located at the vertices of a square whereas the fifth bubble stays at its centroid (FIG.\ref{five-bubbles}). It is observed that the four bubbles located at the vertices decrease simultaneously while, fed by them, the fifth bubble at the center increases until the four disappear, right after it begins decreasing. It shall be called into attention again to the fact that, unlike the LSW theory, the way how the bubbles are settled takes place also in their evolution.   

\begin{figure}[h!]
\centering
\setlength{\unitlength}{1,0mm}
\includegraphics[width=4.25cm,height=4.25cm]{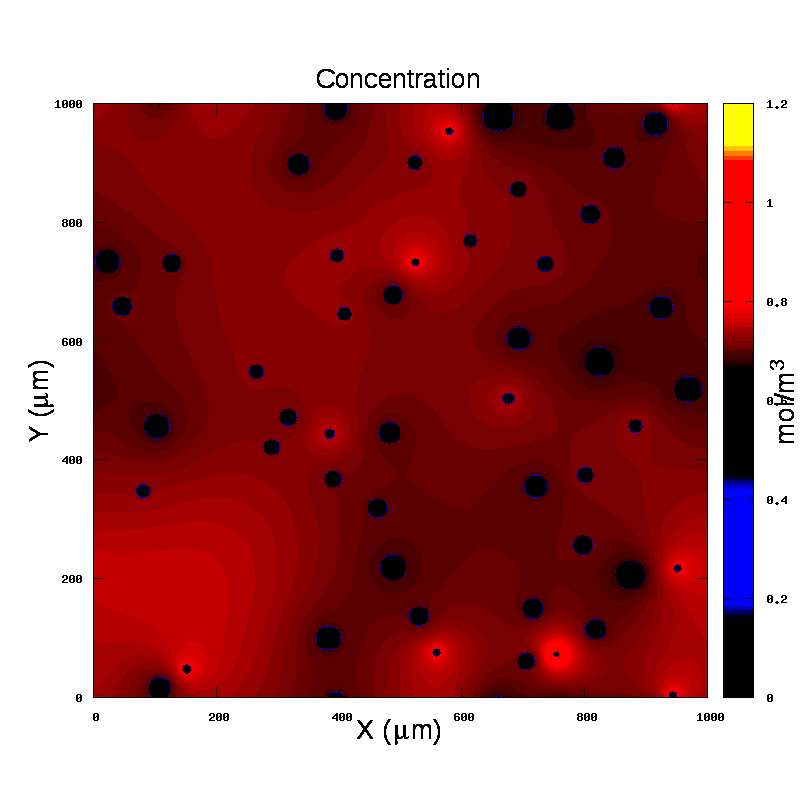}
\includegraphics[width=4.25cm,height=4.25cm]{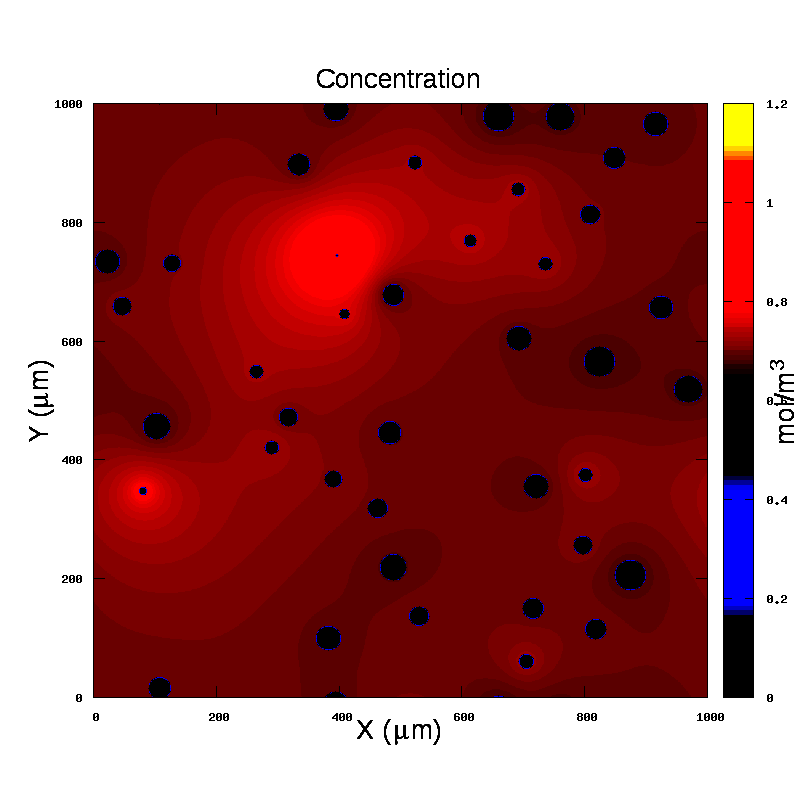}\\
\includegraphics[width=6.0cm,height=4.25cm]{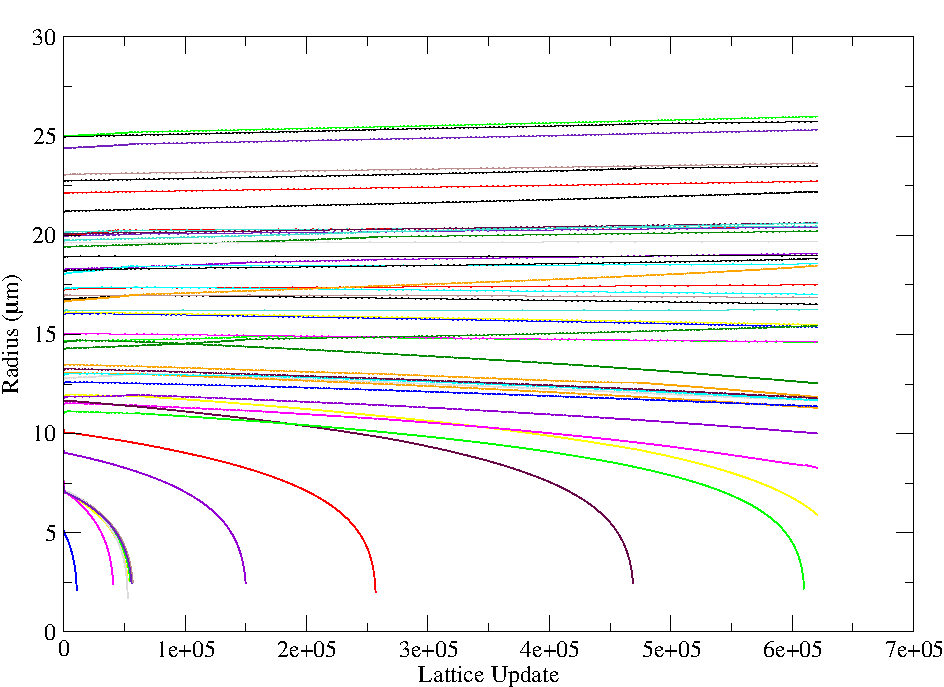}
\caption[]{Simulation of fifty bubbles ($R_{\rm c}=10\m{\rm m}$).}
\label{fifty-bubbles}
\end{figure}

The last and more complex configuration tackled consists of fifty N$_2$ gas bubbles into water -- with initial N$_2$ concentration equivalent to a critical radius given by $R_{\rm c}=10\m{\rm m}$ -- randomly spread  throughout a $1000\m{\rm m}\times1000\m{\rm m}$ lattice, with their initial radii ($R(0)$) aleatory picked up from $5\m{\rm m}$ to $25\m{\rm m}$ (FIG.\ref{fifty-bubbles}). It can be verified from the graph of the bubbles radii time evolution (FIG.\ref{fifty-bubbles}) that even bubbles with initial radii greater than the critical radius ($R_{\rm c}=10\m{\rm m}$) may decrease by feeding other bubbles, unlike to what would be expected from the LSW theory -- where the bubbles with radii smaller than $10\m{\rm m}$ decrease while those with radii greater than $10\m{\rm m}$ increase. LSW theory assumes an infinity volume of the liquid phase, as well as that the distances between bubbles are much greater than their radii, and that the concentration of dissolved gas in the liquid phase remains homogeneous despite its time dependence. Due to all of these assumptions being far from such a system of gas bubbles in a liquid which is the issue in decompression sickness, for that reason it seems to be important to study more deeply the Ostwald ripening phenomenon so as to investigate its contribution to the decompression sickness risks.   

\section{Conclusions and perspectives}
\label{conclusions}
The Ostwald ripening, the phenomenon of gas diffusion among bubbles, which results that larger bubbles are fed by the smaller, are reproduced and described experimentally and by computational simulation. The experiment consisted of air bubbles into a chamber filled by a liquid solution with some human blood-like rheological parameters, density and surface tension. 

The experiment which ran at $25^{\circ}{\rm C}$ under normobaric pressure showed that the number of bubbles ($N(t)$) -- the initial number of bubbles was of order $10^4$ -- decreases in time while the mean bubble radius ($\ov{R}(t)$) increases. It is proposed, analyzing the experimental results, an empirical model for the Ostwald ripening by describing the time evolution of the number of bubbles (\ref{n-time}) and the mean bubble radius (\ref{rbar-time}). It should be stressed that one of the main results shown, even at a constant ambient pressure (at the same diver "depth"), a decreasing number of bubbles in time while the bubbles mean radius increases, consequently, the smaller bubbles disappear whereas the larger (potentially dangerous to the divers, by causing DCS) bubbles grow up, hence this might reveal a contribution of the Ostwald ripening to the decompression sickness risk during and after diving.

The finite element computational simulation, even though at an embryonic stage, has allowed a perception about how critical is the correlation among the distance between bubbles and their time evolution. By way of example, from the case presented above of the three bubbles aligned (FIG.\ref{three-bubbles}), it can be conjectured that the closer a smaller bubble is to a larger bubble the faster it will disappear. Finally, the simulation of fifty bubbles (FIG.\ref{fifty-bubbles}) with radii varying from $5\m{\rm m}$ to $25\m{\rm m}$, within a $1000\m{\rm m}\times1000\m{\rm m}$ lattice, made evident the complexity of a liquid-gas bubbles system when the Ostwald ripening is taken into consideration. Meantime, it became more clear the relevance of the distances between bubbles and their influence to the behaviour of a liquid-gas bubbles system. 

There are many perspectives and challenges to be pursued. From the experimental point of view, and for further computational simulation, it is important to search for mechanisms to suppress or to promote the Ostwald ripening. Besides, the experiment shall be performed at typical human body temperatures, around $36,5\mbox{-}37,5^{\circ}{\rm C}$, for nitrogen bubbles into human plasma, also by adding to the plasma polystyrene microdisks, with diameters about $6\mbox{-}8\m{\rm m}$, simulating the red blood cells. The Ostwald ripening empirical model -- for the number of bubbles (\ref{n-time}) and the mean bubble radius (\ref{rbar-time}) in time -- might be implemented to the Reduced Gradient Bubble Model (RGBM) \cite{diving_physics,rgbm,blood}, first to recreational air diving protocols, in order to obtain the ``new'' risk estimates for various NDLs (no-decompression time limits) and compare them to those of RGBM, as well as to those from other models, namely, ZHL (B\"uhlmann), USN (U.S. Navy) and VPM (Varying Permeability Model).         

\begin{acknowledgments}
O.M.D.C. are deeply indebted to NAUI Worldwide and NAUI Brasil. 

Especially thanks are due to, Dallas Edmiston, Derik Crotts, Angie Cowan, Shannon McCoy, Terrence Tysall, Cheryl Thacker, Peter Oliver, Randy Shaw, Bruce Wienke, Tim O'Leary, Jim Gunderson, Michael Feld, David Ochs, Alvanir ``Jornada'' Silveira de Oliveira, Lilian ``Lica'' Megumi Notomi, M\'arcio Monteiro, Daniel Millikovsky, Sergio Viegas, Irene Demetrescu, Marcos Pedreira Silva, Ant\^onio Carlos Ravasco Ferreira and S\'ilvio da Costa Ferreira Jr.  

Prof. Bruce R. Wienke and Capt. Tim R. O'Leary are strongly acknowledged for invaluable comments and encouragement.

O.M.D.C. dedicates this work to the memory of Randy Shaw (NAUI Worldwide) and Sergio Viegas (DAN Brasil), both friends passed away on July 2017. He also dedicates it to his father (Oswaldo Del Cima, {\it in memoriam}), mother (Victoria M. Del Cima, {\it in memoriam}), daughter (Vittoria) and son (Enzo). 

FAPEMIG, FUNARBE, CNPq and CAPES are acknowledged for invaluable financial help.
\end{acknowledgments}

\bibliographystyle{apsrev4-1}
\bibliography{ensembles}

\end{document}